\documentclass[1p]{elsarticle}

\usepackage{fancyhdr}
\usepackage{booktabs}
\usepackage{ctable}
\usepackage{amsmath}
\usepackage{mathtools}
\usepackage{commath}
\usepackage{graphicx}
\usepackage{nomencl}
\usepackage{commath}
\usepackage{natbib}
\usepackage{multirow}
\usepackage{longtable}
\usepackage{textcomp}
\usepackage{array}
\usepackage{graphics}
\usepackage{tabularx}
\usepackage{color}
\usepackage{geometry}
\usepackage{natbib}
\usepackage{mciteplus}
\usepackage{footnote}
\usepackage{amssymb}
\usepackage{pdflscape}
\usepackage{afterpage}
\usepackage{multirow}
\usepackage[title]{appendix}
\usepackage{subfigure}
\usepackage{caption}
\usepackage{amssymb}
\usepackage{natbib}
\usepackage{mleftright} 
\usepackage{latexsym}
\usepackage{color,soul}

\usepackage{etoolbox}
\renewcommand\nomgroup[1]{%
	\item[\bfseries
	\ifstrequal{#1}{A}{}{%
		\ifstrequal{#1}{G}{Greek Letters}{%
			\ifstrequal{#1}{S}{Subscripts}{}}}%
	]}

\usepackage[version=3]{mhchem} 

\usepackage{lineno,hyperref}

\journal{Physics of Fluids}

\biboptions{numbers,sort&compress}

\begin{document}
\begin{frontmatter}

\title{Computational fluid dynamics investigation of bitumen residues in oil sands tailings transport in an industrial horizontal pipe}


\author[1]{Somasekhara Goud Sontti}
\author[1]{Mohsen Sadeghi}
\author[1]{Kaiyu Zhou}
\address[1]{Department of Chemical and Materials Engineering, University of Alberta, Alberta T6G 1H9, Canada}
\author[2]{Enzu Zheng}
\address[2]{CSIRO Mineral Resources , Clayton, VIC 3168, Australia}


\author[1]{Xuehua Zhang\corref{cor1}}
\ead{Xuehua.Zhang@ualberta.ca}
\cortext[cor1]{Corresponding author}

\begin{abstract}
\noindent 

	Bitumen residues in the oil sand tailings can be a threat to the environment that separating them from tailings before disposal is crucial.  This study establishes an Eulerian-Eulerian computational fluid dynamics (CFD) model for an industrial-scale oil sand tailings pipeline. A comprehensive sensitivity analysis was conducted on the selection of carrier-solid and solid-bitumen drag models. The combination of small and large particle sizes (i.e., 75 \& 700 $\mu$m) and bitumen droplet size (i.e., 400 $\mu$m) provided good agreement with field data in velocity profiles and pressure drop. The validated model was subsequently extended to investigate the influence of the secondary phase (i.e., bitumen droplets and bubbles) on flow characteristics in a tailing pipeline. The investigation covered a range of bitumen droplet size (100-400 $\mu$m), bitumen fraction (0.0025-0.1), bubble size (5-1000 $\mu$m), and bubble fraction (0.0025-0.3) and their influences on the velocity, solids, and bitumen distribution are revealed. For an optimum bubble size of 500 $\mu$m, a maximum recovery of 59\% from the top 50 \% and 83 \% from the top 75 \% of the pipe cross-section was obtained. The present study demonstrates the preferential distribution of bitumen and provides valuable insight on bitumen recovery from an industrial-scale tailings pipeline.
\end{abstract}


\end{frontmatter}

\newpage
\section{Introduction} 
Hydraulic pipeline transport of concentrated slurry flows has a wide range of applications in diverse industries, such as mining, chemistry, oil and waste treatment\cite{pullum2018hydraulic,mohaibes2004aerobic}. Transport of slurry flows by pipeline is considered safe, energy-efficient, and cost-effective. The concentrated slurry flow is a multicomponent system consisting of solid particles, water, and other compounds.  Those fine particles smaller than 44 $\mu$m along with water form non\textendash Newtonian carrier fluid, which typically exhibits shear\textendash thinning behavior\cite{cepuritis2017influence,cruz2019slurry,xiong2022effect}. The presence of coarse particles would form heterogeneous and fully stratified flows due to a low degree of turbulence. It lowers the transport capacity and increases the energy cost simultaneously\cite{shook2015slurry,uzi2018flow,enos1977flow}. In recent decades, many researchers have considerably reported both experimental and numerical studies  of slurry transport in horizontal pipelines. Most of the previous works are concerned with a two\textendash phase slurry system, considers a single and multi\textendash size particle slurry system\cite{pokharel2021impact}.  Regardless, the real world of industrial\textendash scale slurry systems is complex due to the flow's composition and non\textendash Newtonian behavior. \\

In the oil sand industry, pipe transport is used to convey crushed oil sand ores and tailings\cite{crosby2013transporting,kang2021simulation}. Both concentrated slurry flows contain bitumen droplets and trace entrapped gas bubbles in addition to solid particles and water. After extraction of liberated bitumen, the concentrated oil sand tailings composed of a tiny fraction of bitumen residue and high solid contents are transported to the tailing ponds\cite{small2015emissions}. However, the bitumen residue has become a threat to wildlife and the environment \cite{khademi2018provenance,nimana2015energy,dibike2018modelling}. Consequently, it is crucial to separate bitumen from tailings before disposing of them in tailings ponds. The first step to designing a separation technology is understanding the tailings flow and its effective parameters. However, the low bitumen concentration in the slurry and the complexity of the mixture make separation difficult. It is difficult to predict the transport characteristics of slurry flow in large-diameter pipes, especially when there are multiple secondary phase solids and bitumen droplets in the slurry. 

A few experimental works reported a similarly complex multiphase flow system with a gas\textendash liquid\textendash solid flow in a horizontal pipeline. Gillies \textit{et al.}\cite{gillies1997pipeline} experimentally investigated gas\textendash liquid mixtures transport sand in a horizontal pipe in both laminar and turbulent flow regimes. Gas was injected into the loop, and static pressure was measured near the weighed section. They found that the gas injection would increase the solid's transport rate when the flow was turbulent and the axial pressure gradient increased. \citet{scott1971transport} also investigated the experimental study on the transport of solid particles (500 $\mu$m and 100 $\mu$m) by gas\textendash liquid mixtures in horizontal pipes. Experimental observations were reported for different solid concentrations and pipe diameters on the saltation velocity for liquid\textendash solid, bubble, plug, and slug flow regimes. They found that for the larger particles, the effects of bubbles and plugs on the velocity field were insufficient to overcome the forces causing saltation, and there was no significant change in actual saltation velocity.\\

\citet{fukuda1986pressure} also studied pressure drop of an air\textendash water\textendash sand three\textendash phase system in horizontal pipes. Two different flow patterns were observed, i.e., plug and slug flow. They found that pressure drop increased in proportion to the gas velocity. At a constant gas velocity, the differences in pressure drop were due to changes in volumetric particle counteraction. Recently, \citet{zahid2020experimental} experimentally studied the two-phase and three-phase flow behavior in drilling annuli using a high\textendash speed visualization technique. Experimentally, they found that the air\textendash water two\textendash phase flow and gas bubbles were separated by water and  the top of the annulus. A bubbly flow regime was observed for the considered range of operating conditions. However, with an increase in the water and air flow rate, the system pressure was increased. For lower flow rates of air and water, a stratified flow regime was observed with a clear wavy interface on the upper part of the annulus. \citet{kaushal2005effect} experimentally investigated the pressure drop and particle concentration distribution with different combinations of particle size distributions and concentrations. They reported that the particle concentration in the horizontal panel was not correlated with the velocity and overall concentration, and a lower pressure drop was obtained with a broad-grading particle and a lower velocity.\\

To recover residual bitumen before discharging to the tailing ponds, several innovative experimental studies have been conducted to enhance bitumen recovery with microbubble injection in a lab-scale pipeline\cite{wallwork2003bitumen,motamed2020microbubble,zhou2022microbubble,wallwork2004processibility}. Though a vast number of theoretical and experimental studies on liquid\textendash solid flow can be found in the literature, very few studies related to the gas\textendash liquid\textendash solid flow were available, and uncertainties exist in modeling this flow. 
Recently  \citet{motamed2020microbubble} reported experimental work that the optimal bitumen recovery of 50 \% was achieved from the oil sand tailings of 6.68 wt \% sands and 0.2 wt \% bitumen with microbubble injection in a hydrotransport pipeline\cite{motamed2020microbubble}. Furthermore, a higher bitumen recovery of 70 \% from highly concentrated oil sand tailings of 50 wt \% sands was obtained in the following work conducted by \citet{zhou2022microbubble}. Numerous investigations have studied the mechanism of flotation behavior in the presence of microbubbles\cite{xing2017recent,wang2020regulation,zhou2020role}. The interaction between microbubbles and bitumen droplets decreases the system's free energy \cite{li2020bridging}. Due to the longer residence time and high surface-to-volume ratio, microbubbles have a higher probability of collision with bitumen droplets\cite{xing2017recent,gao2021formation}. In addition, microbubbles have faster liquid drainage in the attachment to the bitumen surface\cite{albijanic2010review,qiao2021recent}. Those factors account for enhanced bitumen recovery with microbubble injection. Nevertheless, using experimental techniques, it is challenging to non\textendash intrusively monitor the bitumen-bubble interaction in the turbulent concentrated slurry flow at a high flow rate.\\ 

Simultaneously with the effort to improve experimental results and analytical models, computational fluid dynamics (CFD) is becoming more comprehensive in investigating slurry flows in pipelines. There have been two different approaches to modeling multiphase flows: Eulerian-Eulerian and Eulerian-Lagrangian\cite{messa2021computational,zhang2021optimized}. \citet{kaushal2012cfd} studied the mixture and Eulerian Two\textendash Fluid Model (TFM) to simulate the transport of slurry flow of fine particles up to 50 \% by volume in the pipeline. The Eulerian model contributes a better prediction in both pressure drop and concentration profiles at various overall concentrations and flows velocities compared to the failure of the mixture model in predicting pressure drop with regard to the slurry concentration. \citet{li2018hydrodynamic} numerically simulated the transport of multi\textendash sized slurry through a pipeline employing a steady 3D hydrodynamic model based on the  Kinetic Theory of Granular Flow (KTGF) model. They predict the distributions of velocity and concentration with different particle concentration and sizes, pipe diameter, and slurry velocity.\\ 


Recent works of \citet{li2018effect} and \citet{zhang2021cfd} established the Eulerian multiphase flow model to investigate the effect of particle size of single and multi\textendash sized slurry flows on transport properties under the same conditions, such as flow velocity, wall shear stress, and granular pressure distributions. The presence of fine particles is found to reduce energy consumption by changing the coarse particle’s flow regime.
Shi \textit{et al.}\cite{shi2021impact} carried out simulations on multi\textendash sized slurry flows in the horizontal pipeline under various swirling motions utilizing the Eulerian\textendash Eulerian multiphase model in conjunction with the k\textendash $\omega$ SST scheme. The results suggest that an increased level of swirl results in a higher degree of homogeneity of slurry flows.\\

In general, the mixture model is a low computational effort model than the Eulerian-Eulerian simulation. Our previous work systematically studied the complex multiphase flow system with 8 solid phases, bitumen droplets, and carrier non\textendash Newtonian liquid. However, the mixture model completely ignores the secondary phase interactions \cite{fluent2011ansys}. The equations for the mixture model are relatively similar to those for a single\textendash phase flow but are expressed in terms of the density and velocity of the mixture. As a result,  the secondary phase interactions of bubbles and bitumen droplets are neglected in our previous work due to limitations of the mixture model for high\textendash density ratios\cite{fonty2019mixture,fluent2011ansys}. \citet{ling2003numerical} also reported that for lower flow rates and higher particle concentrations, the mixture model underpredicts the pressure drop.\\


Interestingly, most of the reported research is mainly concerned with the slurry transport for the lab\textendash scale data for Newtonian liquids\cite{li2018hydrodynamic,li2018effect,li2018pressure,zhang2021cfd,shi2021impact,kaushal2012cfd}, while several industrial slurry systems likely exhibit complex non\textendash Newtonian behaviors\cite{zheng2021turbulent,javadi2015laminar,zheng2022dense}. On the other hand, secondary phases like droplets and bubbles play a significant role in industrial scales pipeline transport such as oil sand tailings and mining residuals\cite{elghobashi2019direct,martinez1999breakup,xing2015unified}. In spite of industrial application, most research works completely ignore the fundamental understanding of the droplets and bubble influence in a complex multiphase system phase. The underlying phenomena of droplets and bubbles in a tailings system are imperative and desirable. To the best of our knowledge, there is no published data on slurry systems with droplets and bubbles in the literature that can provide the necessary concentration profiles and pressure drop for industrial applications. Therefore, the main objective of this study is to investigate the effect of the secondary phase interactions on the tailings transport in highly non\textendash Newtonian turbulent flows. The present work would provide an improved design of oil sand tailings pipeline systems, where secondary phases like droplets and bubbles are commonly presented.\\

In this work, we develop a three\textendash dimensional finite volume method (FVM) based on an Eulerian\textendash Eulerian CFD model coupled with KTGF model to investigate the influence of secondary phase droplets and bubbles in an industrial\textendash scale horizontal pipeline tailings system. We systematically conduct the model sensitivity analysis and model validation with the industrial scale field data. The validated CFD model is extended to investigate the effect of  bitumen droplet, bubble size, and secondary phase fraction on flow characteristics. These fundamental understandings can be significantly beneficial for industrial\textendash scale slurry transport systems. This paper is organizedas follows: Sec.\textbf{II} presents the velocity profile and pressure drop of tailing system; Sec.\textbf{III} describes the governing equations of multi-fluid model, turbulence model and non-Newtonian viscosity model; Sec.\textbf{IV} describe the details of our numerical methodology settings, and model validation. Sec.\textbf{V}  we present and discuss the effects of the secondary phase droplet/bubble size and fraction on slurry flow behavior. Sec.\textbf{VI }we conclude our study with some concluding remarks.  
\section{Velocity profile and pressure drop of tailings system } 

\begin{figure}
	\centering
	\includegraphics[width=\textwidth]{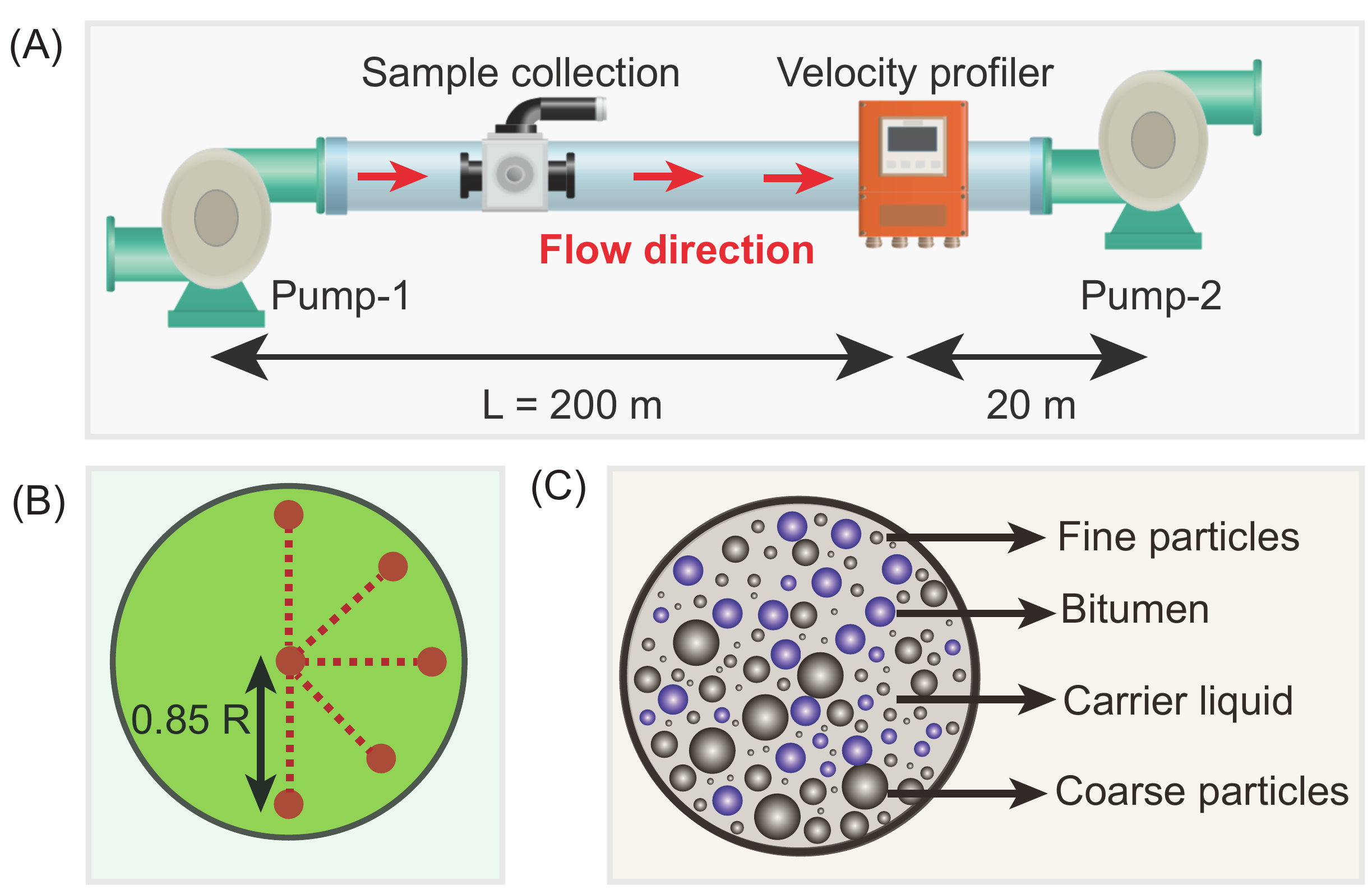}
	\caption{\label{fig:Loop} Schematic representation of (A) tailings hydrotransport pipeline with dimensions, (B) velocity profiler measurement locations, and (C) typical tailing slurry representation. The tailings composition consists of coarse and fine particles with bitumen in a carrier fluid.}
\end{figure}

Fig.\ref{fig:Loop}A shows a schematic of the horizontal pipeline used for field data (i.e., pilot scale hydrotransport pipeline data) collection. The pipe is 220\,m in length and 74\,cm in diameter, with two pumps at the inlet and outlet of the pipe. Several sets of field data were collected on a section of an industrial pipeline for a mining process to validate the CFD model. The samples are collected from the pipe center after the first pump discharge every twelve hours, and the mixture composition is determined using a Dean\textendash Stark apparatus. \citep{dean} To determine the Particle size distribution (PSD) for solid particles sieving method employed \cite{sadeghi2022computational}.

\begin{table}
	\caption{Field data of tailings system used in the CFD model validation.}
	\vspace{0.3cm}
	\label{tab:Fielddata}
	\centering
	\renewcommand{\arraystretch}{1.4}
	\begin{tabular}{|c|c|c|c|c|c|c|}
		\hline
		Case & V $(m/s)$ & \begin{tabular}[c]{@{}c@{}}Solid \\ fraction\end{tabular} & \begin{tabular}[c]{@{}c@{}}Carrier \\ density ($kg/m^{3}$)\end{tabular} & \begin{tabular}[c]{@{}c@{}}Bitumen \\ fraction\end{tabular} & S1 (75 $\mu$m)     & S2 (700 $\mu$m)    \\ \hline
		A     & 5.620   & 0.238                                                   & 1335                                                               & 0.0025                                                      & 0.206 & 0.033 \\ \hline
		B     & 5.353   & 0.230                                                    & 1279                                                               & 0.0033                                                      & 0.199 & 0.031 \\ \hline
		C     & 5.430   & 0.237                                                  & 1329                                                               & 0.0029                                                      & 0.211 & 0.025 \\ \hline
		D     & 5.540   & 0.269                                                    & 1228                                                               & 0.0030                                                      & 0.240 & 0.028 \\ \hline
		E     & 5.764   & 0.223                                                    & 1223                                                               & 0.0029                                                      & 0.192 & 0.030 \\ \hline
		F     & 5.643   & 0.238                                                    & 1332                                                               & 0.0030                                                      & 0.205 & 0.032 \\ \hline
	\end{tabular}
\end{table}

A non-invasive SANDtrac Velocity Profile System (CiDRA) is installed after 200\,m of the pipe inlet and reports the mixture velocity distribution in five locations across the pipe every two seconds. It consists of five sensors attached to the wall of a pipe that track turbulent eddies that cause pressure disturbances and force on the wall. This array senses the dynamic strains applied to the pipe by these forces and converts them to electrical signals to be interpreted, and the velocity values are calculated \citep{sonartrac}. \\

The signals from the profiler are interpreted to the mixture velocity values in different locations. Fig.\,\ref{fig:Loop}B shows the measurement points of the velocity in a cross\textendash sectional view. The device measures the mixture velocities in five locations with a 45\textdegree difference in angle. The radial distance from the pipe center is approximately 0.85R, where R is the pipe radius.\\
The pressure is measured at two locations; at the first pump discharge and the second pump suction every two seconds. The flow rate and mixture density data are also available within a two\textendash second time span. The available data will be used to prepare simulation cases for model validation. Fig.\,\ref{fig:Loop}C shows the schematic of the oil sands tailings flow inside the pipe with its components.\\ 

The selection of a proper time window must be carefully considered after collecting around a million data points and monitoring the fluctuations and significant shifts in flow conditions. Data points are plotted against a 30\textendash minute time window as the initial step. For CFD simulations, a time window is chosen if the oscillations are insignificant and the flow conditions are fairly consistent. After averaging the values over a selected time window, simulation cases are prepared for multiple time windows listed in Table 
\ref{tab:Fielddata}. The particle sizes S1 and S2 represent  Solid 1 and Solid 2 volume fractions of the tailings system.
\section{Mathematical modeling} 

\subsection{Governing equations of multi\textendash fluid model}

Multiphase systems are modeled mathematically as interpenetrating continua in the Eulerian model\cite{sadeghi2022cfd}. The phases are considered separate and treated as interpenetrating continua, which can exchange momentum via the interphase. The interphase exchange coefficient and pressure are responsible for phase coupling. Dissipation of energy, exchange of energy among particles, and interfacial forces are all considered in the model\cite{li2018hydrodynamic,li2018pressure}. \\

\begin{table}
	\centering
	\renewcommand{\arraystretch}{-0.9}
	\begin{tabular}{ m{2cm} m{14cm} }\hline
		
		Continuity \newline\newline & {\begin{subequations}\begin{flalign}
					&\frac{\partial }{\partial t} \left(\rho_q \alpha_q \right)+\nabla \cdot \left (\rho_q \alpha_q \vec{v}_q \right )=0\,, \quad \left(q=l, s, b\right)&\\
					&\alpha_l + \alpha_{s_1} + \alpha_{s_2} + \alpha_b = 1&
				\end{flalign}\label{eq:continuity}\end{subequations}}\\[-4ex]
		\vspace{-4cm}
		Momentum (liquid) \newline\newline \newline\newline \newline\newline \newline & {\begin{subequations}\begin{flalign}
					& \begin{multlined}
						\frac{\partial}{\partial t} \left(\alpha_l \rho_l \vec{v}_l\right) + \nabla \cdot \left(\alpha_l \rho_l \vec{v}_l \otimes \vec{v}_l\right) = - \alpha_l \nabla p + \rho_l \alpha_l \vec{g}  + \nabla \cdot \tau_l  \\ + \vec{F}_{\mathrm{l},s} + \vec{F}_{\mathrm{l},b}  
					\end{multlined}&\\ 
					&\vec{F}_{\mathrm{l},s_i}=\vec{F}_{\mathrm{td},{ls_i}} + \vec{F}_{\mathrm{drag},ls_i} &\\
					&\vec{F}_{\mathrm{drag},ls_i} = \frac{3}{4} C_D \alpha_{si} \rho_l \frac{\norm{\vec{v}_l - \vec{v}_{si}}}{d_p} \left(\vec{v}_l - \vec{v}_{si}\right) &\\
					&\vec{F}_{\mathrm{td},ls_i} = \frac{3}{4} \frac{C_D \mu_{t,l}}{d_p \sigma_{t,l}} \alpha_{si} \norm{\vec{v}_l - \vec{v}_{si}} \left(\frac{\nabla \alpha_{si}}{\alpha_{si}} - \frac{\nabla \alpha_l}{\alpha_l}\right)&\\
					&\vec{F}_{\mathrm{l},b}=\vec{F}_{\mathrm{td},{lb}} + \vec{F}_{\mathrm{drag},lb} &\\
					&\vec{F}_{\mathrm{drag},lb} = \frac{3}{4} C_D \alpha_{b} \rho_l \frac{\norm{\vec{v}_l - \vec{v}_{b}}}{d_b} \left(\vec{v}_l - \vec{v}_{b}\right) &\\
					&\vec{F}_{\mathrm{td},lb} = \frac{3}{4} \frac{C_D \mu_{t,l}}{d_p \sigma_{t,l}} \alpha_{b} \norm{\vec{v}_l - \vec{v}_{b}} \left(\frac{\nabla \alpha_{b}}{\alpha_{b}} - \frac{\nabla \alpha_l}{\alpha_l}\right)&\\
					&
				\end{flalign}\label{eq:momentum_liquid}\end{subequations}}\\[-5ex]
		\vspace{-6cm}
		Momentum ($i$th solid phase) \newline\newline \newline & {\begin{subequations}\begin{flalign}
					& \begin{multlined}
						\frac{\partial}{\partial t} \left(\alpha_{si} \rho_{si} \vec{v}_{si}\right) + \nabla \cdot \left(\alpha_{si} \rho_{si} \vec{v}_{si} \otimes \vec{v}_{si}\right) = - \alpha_{si} \nabla p - \nabla p_{si} + \rho_l \alpha_{si} \vec{g} 
						\\+ \nabla \cdot \tau_{si}  + \vec{F}_{\mathrm{si},l} + \vec{F}_{\mathrm{drag},s_i b} +  \beta_{ij} \left(\vec{v}_{sj} - \vec{v}_{si}\right)
					\end{multlined}&\\
					&\beta_{ij} = \frac{3 \left(1+e_{ij}\right) \left(\frac{\pi}{2} + C_{\mathrm{fr},ij} \frac{\pi^2}{8} \right) \alpha_{si} \rho_{si}\alpha_{sj} \rho_{sj} \left(d_{si} + d_{sj}\right)^2 g_{0,ij}}{2 \pi \left(\rho_{si} d_{si}^3 + \rho_{sj} d_{sj}^3\right)} \norm{\vec{v}_{si} - \vec{v}_{sj}} &\\
					&\tau_{s_i} = \alpha_{s_i} \mu_{s_i} \left(\nabla \vec{v}_{s_i} + \left(\nabla \vec{v}_{s_i}\right)^T\right) +\alpha_{s_i} \left(\lambda_{s_i}-\frac{2}{3} \mu_{s_i}\right) \left(\nabla \cdot \vec{v}_{s_i}\right) \overline{\overline{I}} &\\
					&\vec{F}_{\mathrm{drag},s} = - \vec{F}_{\mathrm{drag},l}; \vec{F}_{\mathrm{td},s} = - \vec{F}_{\mathrm{td},l} &\\
				\end{flalign}\label{eq:momentum_solid_binaryy}\end{subequations}}\\ [-5ex]
		
		\vspace{-4cm}
		Momentum (bitumen) \newline\newline \newline & {\begin{subequations}\begin{flalign}
					& \begin{multlined}
						\frac{\partial}{\partial t} \left(\alpha_{b} \rho_{b} \vec{v}_{b}\right) + \nabla \cdot \left(\alpha_{b} \rho_{b} \vec{v}_{b} \otimes \vec{v}_{b}\right) = - \alpha_{b} \nabla p + \rho_l \alpha_{b} \vec{g} 
						\\+ \nabla \cdot \tau_{b}  + \vec{F}_{\mathrm{b},l} + \vec{F}_{\mathrm{b},si} \end{multlined}&\\
					&\vec{F}_{\mathrm{b},l} = -\vec{F}_{\mathrm{l},b} ;  \vec{F}_{\mathrm{b},si} = -\vec{F}_{\mathrm{si},b}&\\ 
				\end{flalign}\label{eq:momentum_solid_binary}\end{subequations}}\\
		
		\hline\end{tabular}\caption{Momentum equations\cite{fluent2011ansys,sadeghi2022cfd}.}\label{tab:CFD_model_equations_Binary}
\end{table}

\begin{table}
	\centering
	\abovedisplayskip=0pt
	\belowdisplayskip=0pt
	\renewcommand{\arraystretch}{0.1}
	\begin{tabular}{ m{2cm} m{14cm} }\hline
		
		Granular kinetic theory ($i$th solid phase) \newline\newline \newline\newline \newline\newline \newline\newline \newline\newline \newline\newline \newline \newline\newline & {\begin{subequations}\begin{flalign}
					& \lambda_{s_i} = \frac{4}{3} \alpha^2_{si} \rho_{si} d_{si} g_{0,ii} \left(1+e_{ij}\right) \left(\frac{\Theta_{si}}{\pi}\right)^{1/2} &\\
					& g_{0,ii} = \left[1-\left(\sum_{i=1}^2 \alpha_{si}/\alpha_{s,\mathrm{max}}\right)^{1/3}\right]^{-1} + \frac{d_{si}}{2} \sum_{i=1}^2 \frac{\alpha_{si}}{d_{si}} &\\
					& g_{0,ij} = \frac{d_{si} g_{0.ii} + d_{sj} g_{0,jj}}{d_{si} + d_{sj}} &\\
					& \Theta_{si} = \frac{1}{3} \norm{\vec{v}_{si}^{\,\prime}}^2 &\\
					& 0 = \left(-p_{si} \overline{\overline{I}} + \overline{\overline{\tau_{si}}}\right):\nabla \vec{v}_{si} -\gamma_{\Theta_{si}} +\phi_{li}&\\
					&\gamma_{\Theta_s}=\frac{12\left(1-e^2_{ii}\right) g_{0,ii}}{d_{si} \pi^{1/2}}\rho_{si}\alpha^2_{si} \Theta^{3/2}_{si} &\\
					& \phi_{li}=-3K_{li}\Theta_i &\\
					& p_{si}  = \alpha_{si} \rho_{si} \Theta_{si} \left[1 + 2 \sum_{j=1}^2 \left(\frac{d_{si}+d_{sj}}{2d_{si}}\right)^3 \left(1+e_{ij}\right) \alpha_{sj} g_{0,ij} \right]&\\
					& \mu_{si}  = \mu_{si,\mathrm{col}} + \mu_{si,\mathrm{kin}} + \mu_{si,\mathrm{fr}}&\\
					& \mu_{si,\mathrm{col}} = \frac{4}{5} \alpha_{si} \rho_s d_{si} g_{0,\mathrm{ii}} \left(1+e_{ij}\right) \left(\frac{\Theta_{si}}{\pi}\right)^{1/2}\alpha_{si}&\\
					& \mu_{si,\mathrm{kin}} = \frac{10 \rho_{si} d_{si} \left(\Theta_{si} \pi\right)^{1/2}}{96 \alpha_{si} \left(1+e_{ij}\right) g_{0,\mathrm{ii}}} \left[1 + \frac{4}{5} g_{0,\mathrm{ii}} \alpha_{si} \left(1+e_{ii}\right)\right]^2\alpha_{si}&\\
					& \mu_{si,\mathrm{fr}} = \frac{p_{si} \sin \varphi_{si}}{2 I_{2D}^{1/2}} &
				\end{flalign}\label{eq:granular_kinetic_theory_binary}\end{subequations}}\\

		\hline\end{tabular}\caption{Eqations from granular kinetic theory\cite{fluent2011ansys,sadeghi2022cfd,li2018hydrodynamic}.}\label{tab:CFD_ktgf}
\end{table}

Table\,\ref{tab:CFD_model_equations_Binary} lists the governing equation of the mass and momentum balance for the phases. Eq.\,\eqref{eq:continuity} shows the continuity equation for all of the phases. Eq.\,\eqref{eq:momentum_liquid} shows the momentum balance for the liquid (carrier fluid), which is the primary phase in this study. In this equation, $\vec{F}_{\mathrm{l},s}$ and $\vec{F}_{\mathrm{l},b}$ refer to the interphase forces between the liquid with the solids and bitumen phases, respectively. For the interactions between the liquid and solid phases, the drag force (Eq.\,\eqref{eq:momentum_liquid}\.c), virtual mass (Eq.\,\eqref{eq:momentum_liquid}\.d), and turbulent dispersion force (Eq.\,\eqref{eq:momentum_liquid}\.e) are included in the model. The drag force arises from the difference between the velocities of the primary and secondary phases in the flow direction. The drag force has been proven to be an essential force in the modeling of multiphase slurry flows. Gidaspow \textit{et al.}\cite{gidaspow1991hyd} drag model has been extensively used by other researchers in the literature and proven to accurately describe the drag force between the solid and liquid phase \cite{sadeghi2022cfd, li2018hydrodynamic,li2018pressure}. In a turbulent slurry flow, the interactions between the turbulent eddies and secondary phases resulting in the turbulent dispersion force can significantly influence the flow behavior and should be included in the model\cite{burns2004favre,ting2019comparative,antaya}. To this end, the model introduced by \citet{burns2004favre} is implemented to account for the turbulent dispersion force between the carrier and solid particles.\\

For the interactions between the carrier fluid and bitumen droplets, the drag (Eq.\,\eqref{eq:momentum_liquid}f) and turbulent dispersion (Eq.\,\eqref{eq:momentum_liquid}g) forces have been included in the model similar to carrier\textendash solids interactions. The drag model used for the carrier\textendash bitumen is the Symmetric model\citep{fluent2011ansys}, and \citet{burns2004favre} model has been implemented for the turbulent dispersion force. The Gidaspow \textit{et al.}\cite{gidaspow1991hyd} drag force has been included to capture the interphase force between the solid phases. And for the bitumen and solid phases, the Symmetric drag model has been used. 

\subsection{Governing equations of turbulence model}
A mixture turbulence $k$-$\epsilon$ model based on Reynolds-averaged Navier-Stokes (RANS) equations is used to capture the turbulent ice slurry flow \cite{li2018hydrodynamic,liu2022effect}. The $k$-$\epsilon$ model equations describing this model are listed in Table \ref{tab:Turbulence}
\begin{table}
	\centering
	\abovedisplayskip=0pt
	\belowdisplayskip=0pt
	\begin{tabular}{ m{3cm} m{13cm} }\hline
		
		The $k$ 
		equation\newline\newline \newline\newline & {\begin{subequations}\begin{flalign}
					&\frac{\partial}{\partial t}\left(\rho_m k\right)+\nabla\left(\rho_m \vec{v}_m k\right)=\nabla \cdot\left(\frac{\mu_{t, m}}{\sigma_k} \nabla k\right)+G_{k, m}-\rho_m \varepsilon&
				\end{flalign}\label{eq:kequation}\end{subequations}}\\[-5ex]
		
		The $\epsilon$
		equation \newline \newline\newline  & {\begin{subequations}\begin{flalign}
					& 
					\frac{\partial}{\partial t}\left(\rho_m \varepsilon\right)+\nabla\left(\rho_m \vec{v}_m \varepsilon\right)=\nabla \cdot\left(\frac{\mu_{t m}}{\sigma_{\varepsilon}} \nabla \varepsilon\right)+\frac{\varepsilon}{k}\left(C_{1 \varepsilon} G_{k, m}-C_{2 \varepsilon} \rho_m \varepsilon\right) 
					&
				\end{flalign}\label{eq:epislon}\end{subequations}}\\[-5ex]
		
		Mixture density
		\newline \newline\newline  & {\begin{subequations}\begin{flalign}
					& 
					\rho_m=\sum_{q=1}^n \alpha_q \rho_q
					&
				\end{flalign}\label{eq:density}\end{subequations}}\\[-5ex]
		
		Mixture velocity \newline \newline\newline  & {\begin{subequations}\begin{flalign}
					&  \vec{v}_m=\left(\sum_{q=1}^n \alpha_q \rho_q \vec{v}_q\right) /\left(\sum_{i=q}^n \alpha_q \rho_q\right)
					&
				\end{flalign}\label{eq:velocity}\end{subequations}}\\[-5ex]
		
		Turbulent\\ viscosity \newline \newline\newline  & {\begin{subequations}\begin{flalign}
					&  \mu_{t, m}=\rho_m C_\mu \frac{k^2}{\varepsilon}
					&
				\end{flalign}\label{eq:viscosity}\end{subequations}}\\[-1ex]
		
		Standard\\ constants \newline \newline\newline  & {\begin{subequations}\begin{flalign}
					&  C_{1 \varepsilon}=1.44, C_{2 \varepsilon}=1.92, C_\mu=0.09, \sigma_k=1.0, \sigma_{\varepsilon}=1.3
					&
				\end{flalign}\label{eq:CCC}\end{subequations}}\\[-5ex]
		
		\hline\end{tabular}\caption{Stanadrd $k$-$\epsilon$ mixture turbulence model.\cite{li2018hydrodynamic}}\label{tab:Turbulence}
\end{table}

\subsection{Casson viscosity model }

According to \citet{rheology}, the suspension of fine sand particles in water with a concentration in the range of 10--40\,wt\% follows the Casson rheological model. As the mass fraction of the fine particles in this study fall into the mentioned range, the non-Newtonian behavior of the carrier model can be modeled via the Casson model. The equation for this model is expressed via Eq.\,\eqref{eqn:Cassonmodel}, where $\mu_c$ is the Casson viscosity. 

\begin{equation} \label{eqn:Cassonmodel}
	\tau^{1/2}=\tau_{\mathrm{y}}^{1/2}+\mu_{\mathrm{c}}^{1/2} \dot{\gamma}^{1/2}
\end{equation}

\section{Implementation and validation of CFD models}
\subsection{Computational model and solver settings} 
In this work, a three\textendash dimensional circular pipeline with the inner diameter of D = 0.74 m and length of Z = 105 m is considered for the numerical investigation based on the industrial scale pipeline conditions as shown in Fig.\ref{fig:Mesh}A. Based on the computed velocity profiles along the slurry pipeline, it is confirmed that the flow is fully developed. This study employs an unsteady state Eulerian multiphase model in which different phases are conceptualized as interpenetrating continuous systems. To describe particle interactions, granular kinetic theory is used.  All phases share a single pressure, and each phase solves its corresponding conservation equations for mass, momentum, and energy. All phases are coupled by pressure and interphase exchange coefficients. In this method, volume fractions of continuous and dispersed phases are assumed to be continuous functions of space and time, and their sum is equal to one. Also, interphase exchange coefficients are used to model all secondary phase interactions. For secondary phase interactions, the Eulerian\textendash Eulerian method is more comprehensive and robust from a computational perspective. The current study uses the Eulerian\textendash Eulerian  method. \\

Table. \ref{tab:CFD_model_equations_Binary} and Table. \ref{tab:CFD_ktgf} lists the conservation equation for the Eulerian\textendash Eulerian multi\textendash fluid model (MFM) model with KTGF. A finite volume method (FVM) based commercial
software \textit{ANSYS Fluent} solver 2020 R2 is used to solve all unsteady state equations\cite{fluent2011ansys}. The details of the solver settings and schemes are presented in Table \ref{tab:Solversettings}. At the pipe inlet, each phase's velocity and volume concentration is assumed to be uniform. An outlet boundary condition equal to atmospheric pressure is selected as the outlet boundary condition. At the wall, the liquid phase velocity is set to zero, corresponding to the no\textendash slip condition. Turbulence intensity and turbulent viscosity ratio of all phases are set to the values 5\% and 10, respectively\cite{li2018hydrodynamic}. The details of all the fluid properties and KTGF model parameters are listed in Table \ref{tab:Materialparameters}.\\

\begin{table}
	\caption{List of different models and solver settings details of the multi\textendash fluid model.}
	\vspace{0.1cm}
	\label{tab:Solversettings}
	\resizebox{\textwidth}{!}{%
		\begin{tabular}{ll}
			\specialrule{.1em}{.05em}{.05em} 
			\multicolumn{1}{c}{Model}                      & \multicolumn{1}{c}{Scheme} \\ \specialrule{.1em}{.05em}{.05em} 
			Multiphase model                                 & Eulerian           \\
			Turbulence model                                    & $k$\textendash $\epsilon$ standard \cite{li2018hydrodynamic}               \\
			Turbulent dispersion                      & \citet{burns2004favre}  \\
			Turbulence Multiphase                       & Mixture \cite{sadeghi2022cfd} \\
			Carrier\textendash solid drag                         & Gidaspow \cite{gidaspow1991hyd,sadeghi2022cfd}\\
			Carrier\textendash bitumen drag                       & Symmetric \cite{fluent2011ansys,sen2016cfd}\\
			Solid\textendash bitumen drag                          & Symmetric  \cite{fluent2011ansys,sen2016cfd}\\
			Solid\textendash Solid drag                             & Gidaspow \cite{gidaspow1991hyd,sadeghi2022cfd}\\
			Pressure\textendash velocity coupling                      & Phase coupled SIMPLE\\
			Pressure                                        & PRESTO\\
			Momentum \& volume fraction                     & Second order upwind \\
			Turbulent kinetic energy \& dissipation rate    & Second order upwind \\
			Transient formulation                           & Second order upwind\\
			Carrier fluid shear condition                    & No\textendash slip \\
			Bitumen shear condition                         & No\textendash slip \\
			Carrier viscosity                                & Casson viscosity model \cite{rheology} \\
			Time step                                       & 0.01 s\\
			Number of time steps                            & 20,000\\
			
			\specialrule{.1em}{.05em}{.05em} 
		\end{tabular}%
	}
\end{table}
For solving momentum equations, the second\textendash order upwind method is used, while for solving volume fraction, turbulence transport, and other equations, a second\textendash order upwind method is used, with a pressure relaxation factor of 0.3, a momentum relaxation factor of 0.7, and a volume fraction relaxation factor of 0.4. In the present study, density, body forces, granular temperature, turbulent kinetic energy, turbulent dissipation rate, and turbulent viscosity relaxation factors are concurrently maintained at their default values of 1, 1, 0.2, 0.8, 0.8, and 1, respectively. For each scaled residual component, a convergence criterion of $10^{-4}$ is defined. The gravitational acceleration $g = -9.8$ $m^{2}/s$ is considered in Y\textendash direction. All the simulations are performed in high\textendash performance computing (HPC) facility at Compute Canada Ceder cluster with 44 CPUs, and the simulations are solved for 20,000\textendash time steps for the period of 200 $s$ flow time. After reaching 100 $s$ most of the simulation reached a steady state. However, all simulations are run for 200 $s$ to obtain accurate and reliable data.   \\

\begin{table}
\centering
	\caption{Material parameters and boundary conditions used in CFD simulations.}
	\vspace{0.3cm}
	\label{tab:Materialparameters}
	\resizebox{0.60 \textwidth}{!}{%
		\begin{tabular}{ll}
			\specialrule{.1em}{.05em}{.05em} 
			\multicolumn{1}{c}{Parameter}                      & \multicolumn{1}{c}{value} \\ \specialrule{.1em}{.05em}{.05em} 
			Pipe diameter, m                                 & 0.74           \\
			Pipe length, m                                   & 105           \\
			Particle diameter, $\mu$m                        & 75 \& 700           \\
			Density of the particle, $kg/m^{3}$              & 2650           \\
			Carrier density, $kg/m^{3}$                       & 1335           \\
			Bitumen viscosity, $Pa.s$                         & 20           \\
			Casson  viscosity $\mu_{c}$, $Pa^{1/2} s^{1/2}$                         & 0.0035 \cite{rheology}          \\
			Yield stress, $\tau_{y}$, $Pa$                         & 0.0016  \cite{rheology}         \\
			Fraction packing limit                                 & 0.60  \cite{fluent2011ansys}         \\
			Angle of internal friction                                & 30 \cite{fluent2011ansys}          \\
			Particle\textendash particle restitution coefficient   & 0.90 \cite{li2018hydrodynamic,zhang2021influence}        \\
			Particle\textendash wall specularity coefficient                                & 0.20 \cite{liu2021computational}          \\
			
			\specialrule{.1em}{.05em}{.05em} 
		\end{tabular}%
	}
\end{table}

\subsection{Gird independence study}
\begin{figure}
	\centering
	\includegraphics[width=0.85\textwidth]{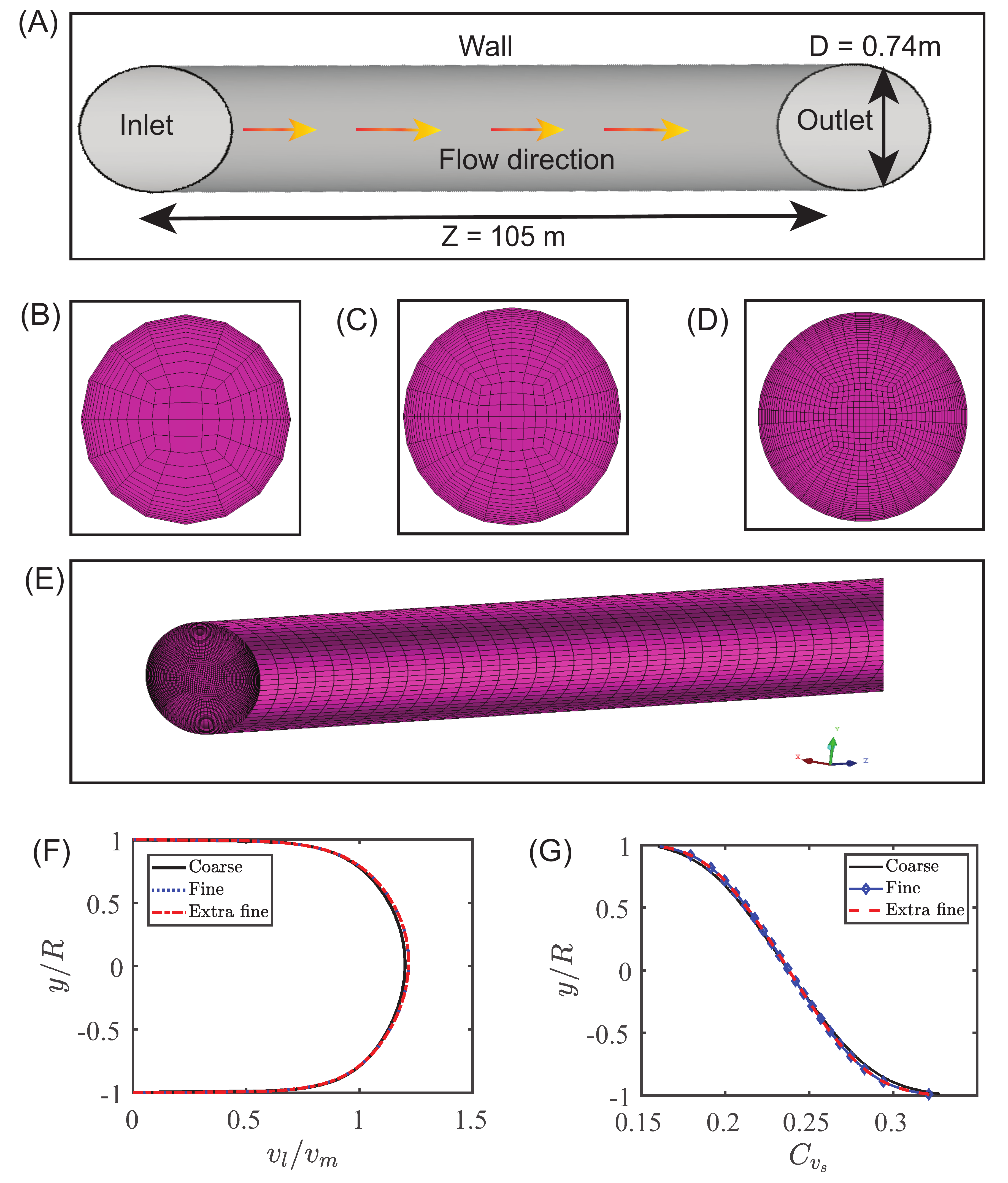}
	\caption{\label{fig:Mesh} (A) Computational domain with dimensions and boundary conditions. Cross\textendash sectional view of mesh for different mesh structures (B) coarse,  (C) fine, (D) extra fine, and  (E) grid structure along the length of the pipeline. Comparison of (F) velocity profile and (G) solid volume fraction profiles for different meshes.}
\end{figure}

At first grid independence study is conducted  to understand the mesh density on flow characteristics, as depicted in Fig.\ref{fig:Mesh}B\textendash D. Three different meshes like s coarse, fine, and extra fine meshes, are examined to ensure good quality computations and convergence of the models. Number of nodes for coarse, fine and extra fine meshes are 1,68,682, 3,39,500 and 5,27,253 respectively.  To ensure the accuracy of a computational model and the near\textendash the\textendash wall effect, 30  boundary layers are considered. The 3D computational structured mesh is portrayed in Fig.\ref{fig:Mesh}E. The velocity profiles along the vertical reference line are analyzed at Z = 100 m for all the cases. Fig.\ref{fig:Mesh}F and G results demonstrated that fine and extra fine mesh results are almost identical for velocity profile and also solids concentration profiles. The results demonstrated that the considered grid and number of nodes are sufficient to accurately capture the flow physics.          
\subsection{Drag models and particle size sensitivity analysis}
The momentum exchange between the two dispersed phases viz, droplets and solid phases, have to be taken into account for CFD simulation of three\textendash phase  and four\textendash phase flows since the droplets/bubbles tend to follow in the slurry system like a fluid phase. The selection of appropriate drag models is also essential for multiphase modeling\cite{liu2022effect}. To study the sensitivity of drag models between carrier\textendash bitumen and solid\textendash bitumen, drag models are carefully studied. In the open literature, different drag models are available such as schiller\textendash naumann \cite{schiller}, morsi\textendash alexander \cite{morsi_alexander_1972}, Symmetric \cite{fluent2011ansys}, Grace \cite{clift2005bubbles}, Tomiyama \cite{tomiyama}, and ishii\textendash Zuber \cite{ishii}. \\   
\begin{figure}
	\centering
	\includegraphics[width=\textwidth]{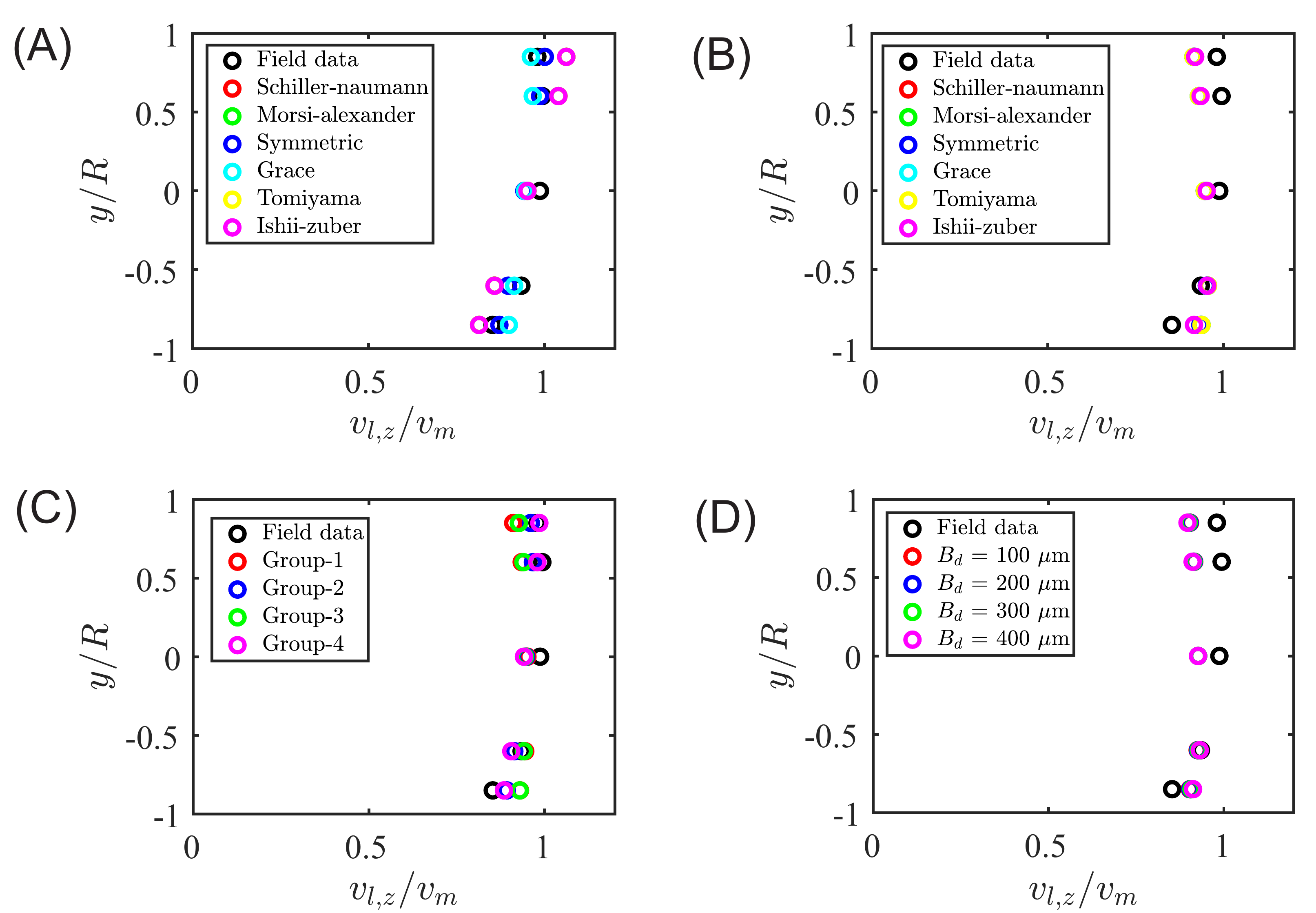}
	\caption{\label{fig:sensivity} Comparison of a velocity profile for different  (A) carrier\textendash bitumen drag models, (B) solid\textendash bitumen drag models, and (C) combination of particle sizes. Group\textendash 1: 200 \& 500 $\mu$m, Group\textendash 2: 200 \& 700 $\mu$m, Group\textendash 3: 75 \& 500 $\mu$m, and Group\textendash 4: 75 \& 700 $\mu$m. (D) Bitumen droplet size. All the properties are considered for case\textendash A, from Table \ref{tab:Fielddata}. }
\end{figure}

Fig.\ref{fig:sensivity}A, demonstrates the velocity profile data comparison with field data for several carrier\textendash bitumen drag models. In other words, the drag model between the primary and secondary phases. All the drag models showed similar trends with the field data points. However, Symmetric drag models showed better prediction compared to other models, with a maximum deviation of 2.71 \%  in terms of average velocity deviation. All the other drag model predictions are distinct at the bottom of the pipe and the top part of the pipe. Notably, the average velocity deviation is two\textendash fold higher than the Symmetric drag model.\\

\begin{table}
	\caption{A comparison of velocity and pressure error percentages for different drag models.}
	\vspace{0.3cm}
	\label{tab:Pressuredrop}
	\centering
	\renewcommand{\arraystretch}{1.4}
	\resizebox{ \textwidth}{!}{%
		\begin{tabular}{|c|cc|cc|}
			\hline
			\multirow{2}{*}{Model} & \multicolumn{2}{c|}{\begin{tabular}[c]{@{}c@{}}Average velocity\\  error \%\end{tabular}} & \multicolumn{2}{c|}{\begin{tabular}[c]{@{}c@{}}Pressure drop \\ error \%\end{tabular}} \\ \cline{2-5} 
			& \multicolumn{1}{c|}{Carrier\textendash bitumen}                   & Solid\textendash bitumen                   & \multicolumn{1}{c|}{Carrier\textendash bitumen}                  & Solid\textendash bitumen                 \\ \hline
			Schiller\textendash nauman \cite{schiller}        & \multicolumn{1}{c|}{5.84}            & 5.78          & \multicolumn{1}{c|}{1.28}            & 4.69          \\ \hline
			Moris\textendash alexander \cite{morsi_alexander_1972}       & \multicolumn{1}{c|}{5.85}            & 5.80          & \multicolumn{1}{c|}{1.26}            & 3.77          \\ \hline
			Symmetric \cite{fluent2011ansys}              & \multicolumn{1}{c|}{2.71}            & 5.62          & \multicolumn{1}{c|}{4.86}            & 4.20          \\ \hline
			Grace \cite{clift2005bubbles}                 & \multicolumn{1}{c|}{5.84}            & 5.81          & \multicolumn{1}{c|}{1.16}            & 3.40          \\ \hline
			Tomiyama \cite{tomiyama}           & \multicolumn{1}{c|}{5.84}            & 5.79          & \multicolumn{1}{c|}{1.28}            & 2.83          \\ \hline
			Ishi\textendash zuber \cite{ishii}            & \multicolumn{1}{c|}{5.58}            & 5.14          & \multicolumn{1}{c|}{1.25}            & 9.20          \\ \hline
		\end{tabular}
	}
\end{table}
Since the two dispersed phases are assumed to be continua in our system, it is necessary to model the drag force between the solid particles and droplets/bubbles in the same way as the primary\textendash secondary phases. The drag model between the solid\textendash bitumen (i.e., drag models between secondary phases) is comprehensively studied as shown in Fig.\ref{fig:sensivity}B. Even though all the drag models showed similar agreement with the field data, the results are further analyzed based on the pressure drop and compared with field data as listed in Table \ref{tab:Pressuredrop}. The Symmetric drag model exhibited trustworthy prediction in terms of velocity profile agreement and pressure drop. Therefore, the Symmetric drag model is considered between the carrier\textendash bitumen and solid\textendash bitumen in this study. The most popular Gidaspow \cite{gidaspow1991hyd} drag model is used between the carrrier\textendash solid and solid\textendash solid phase. A recent study of \citet{sadeghi2022cfd} also successfully demonstrated the applicability of Gidaspow \cite{gidaspow1991hyd} drag model prediction in a slurry system.\\

\begin{table}
	\caption{A comparison of velocity and pressure error percentages for bitumen droplet and particle size combinations.}
	\vspace{0.3cm}
	\label{tab:Bitumendroperror}
	\centering
	\renewcommand{\arraystretch}{1.4}
	\begin{tabular}{|c|c|c|c|}
		\hline
		Study                                  & Size ($\mu$m) & \begin{tabular}[c]{@{}c@{}}Average velocity \\ error \%\end{tabular} & \begin{tabular}[c]{@{}c@{}}Pressure drop \\ error \%\end{tabular} \\ \hline
		\multirow{4}{*}{Particle combination} & Group\textendash 1: 200 \& 500  & 5.36                                                                 & 4.86                                                              \\ \cline{2-4} 
		& Group\textendash 2: 200 \& 700  & 3.11                                                                 & 4.65                                                              \\ \cline{2-4} 
		& Group\textendash 3: 75 \& 500   & 4.97                                                                 & 1.30                                                              \\ \cline{2-4} 
		& Group\textendash 4: 75 \& 700   & 2.70                                                                 & 0.61                                                              \\ \hline
		\multirow{4}{*}{Bitumen  droplet}     & 100         & 5.81                                                                 & 1.65                                                              \\ \cline{2-4} 
		& 200         & 5.81                                                                 & 1.89                                                              \\ \cline{2-4} 
		& 300         & 5.80                                                                 & 2.24                                                              \\ \cline{2-4} 
		& 400         & 5.78                                                                 & 2.69                                                              \\ \hline
		
	\end{tabular}
\end{table}

Furthermore, the effect of particle size combination and bitumen droplet is also comprehensively investigated. Simulating the whole particle sizes is challenging due to computational time and convergence issues with Eulerian-Eulerian multifluid models. To simplify the computational model, two solid particles are considered by covering the whole range of particle sizes. Small and larger particles are considered in different combinations to cover the full PSD range. Fig.\ref{fig:sensivity}C shows the velocity profile field data agreement with a range of particle size combinations. The CFD prediction revealed that Group\textendash4 in combination with smaller particle size 75 $\mu$m and coarse particle size 700 $\mu$m showed excellent agreement with field data measurements, and the corresponding pressure drop is also found to be good in agreement. The considered particle size combination with smaller particle size 75 $\mu$m and coarse particle size 700 $\mu$m represents the general oil sand tailing system. These results also indicate small and large particle combinations are more likely to describe the tailings compositions. Therefore, based on the velocity profile and pressure drop data (Table \ref{tab:Bitumendroperror} comparison with the field data, Group\textendash4 particle size combination (i.e., Solid 1\textendash 75 $\mu$m and Solid 2\textendash 700  $\mu$m) is considered for further investigations. For simplicity, Solid 1 and Solid 2 are referred to as S1 and S2.  The influence of bitumen droplet size is also investigated for different ranges from  100 $\mu$m to 400$\mu$m based on the literature data \cite{malysa1999method,malysa1999method2}. Fig.\ref{fig:sensivity}D revealed that trends are identical for all the considered cases. Therefore, on the basis of a comprehensive analysis of velocity profile  and pressure drop agreement with the field data, the bitumen droplet size is chosen as 400 $\mu$m in the present study.   \\

\subsection{Model validation with field data}

\begin{figure}
	\centering
	\includegraphics[width=\textwidth]{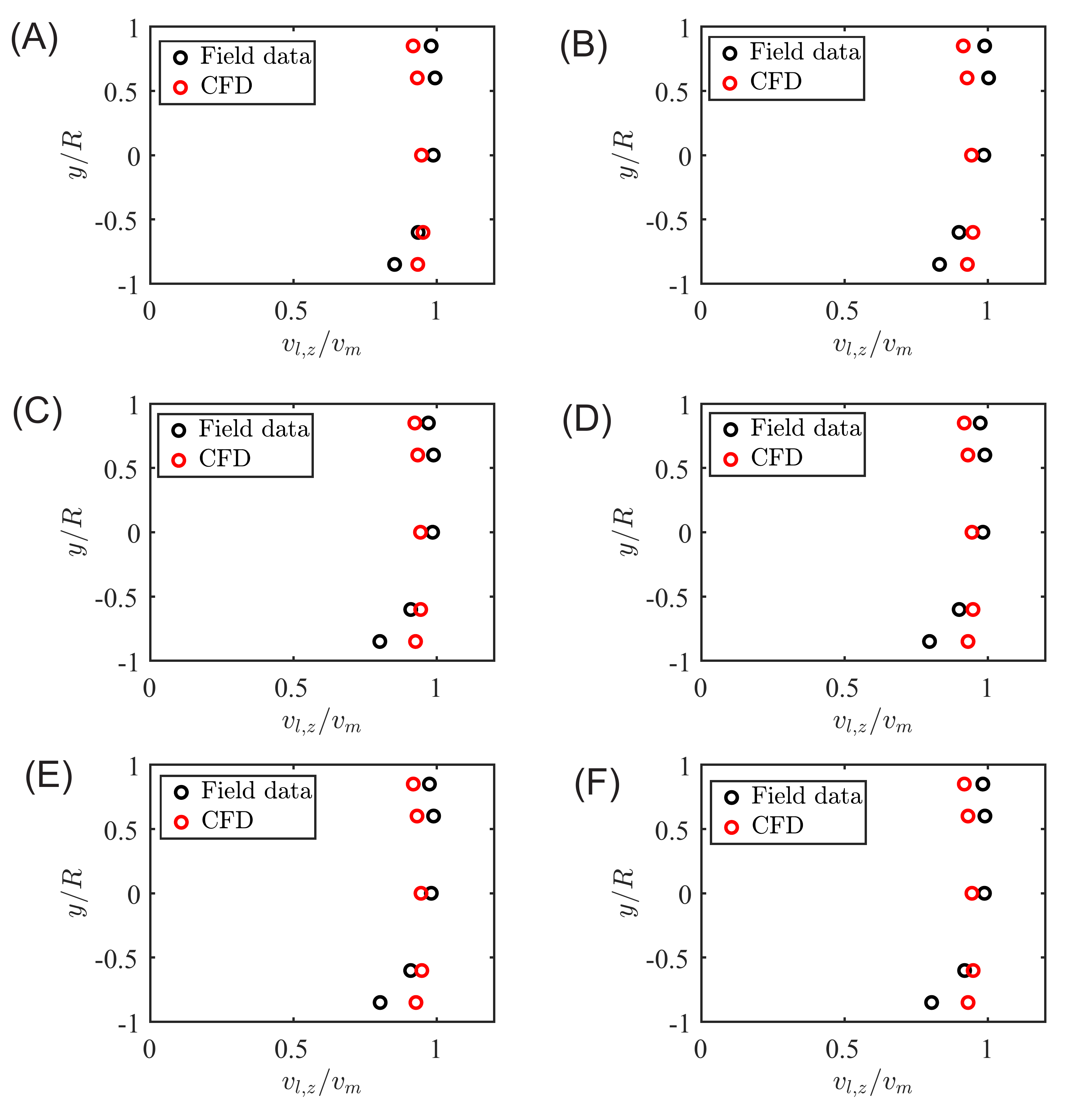}
	\caption{\label{fig:ValidationCFD} Comparison of velocity distribution field data with CFD model results for different cases of (A)\textendash (E). The red circles represent the CFD predictions, and the black color represents the field data. Detailed data on all the fluid properties and flow conditions for pressure drop validation are provided in Table \ref{tab:Fielddata}.}
\end{figure}

To demonstrate the accuracy of the developed CFD model, the model forecasts are examined with 6 sets of field data of velocity distribution and pressure gradient. Fig.\ref{fig:ValidationCFD}A\textendash F provides an overview of the comparison between the CFD\textendash predicted velocity distribution and the measured ones at the field for six different sets. The comprehensive CFD model is established by carefully considering all the sensitivity investigations and model parameters discussed in the previous section. The CFD model predictions are found to be in excellent accordance with the carrier fluid velocity field data. For all the cases, the maximum average velocity error is lower than 5 \%. The maximum error is found near the bottom wall for all the cases. This might be due to the accumulation/dynamics of coarse solid particles at the bottom wall. On the other hand, measuring the accurate filed data is also challenging due to bed formation at the bottom wall. However, the CFD model agreements showed excellent agreement with other data points.\\
\begin{figure}
	\centering
	\includegraphics[width=\textwidth]{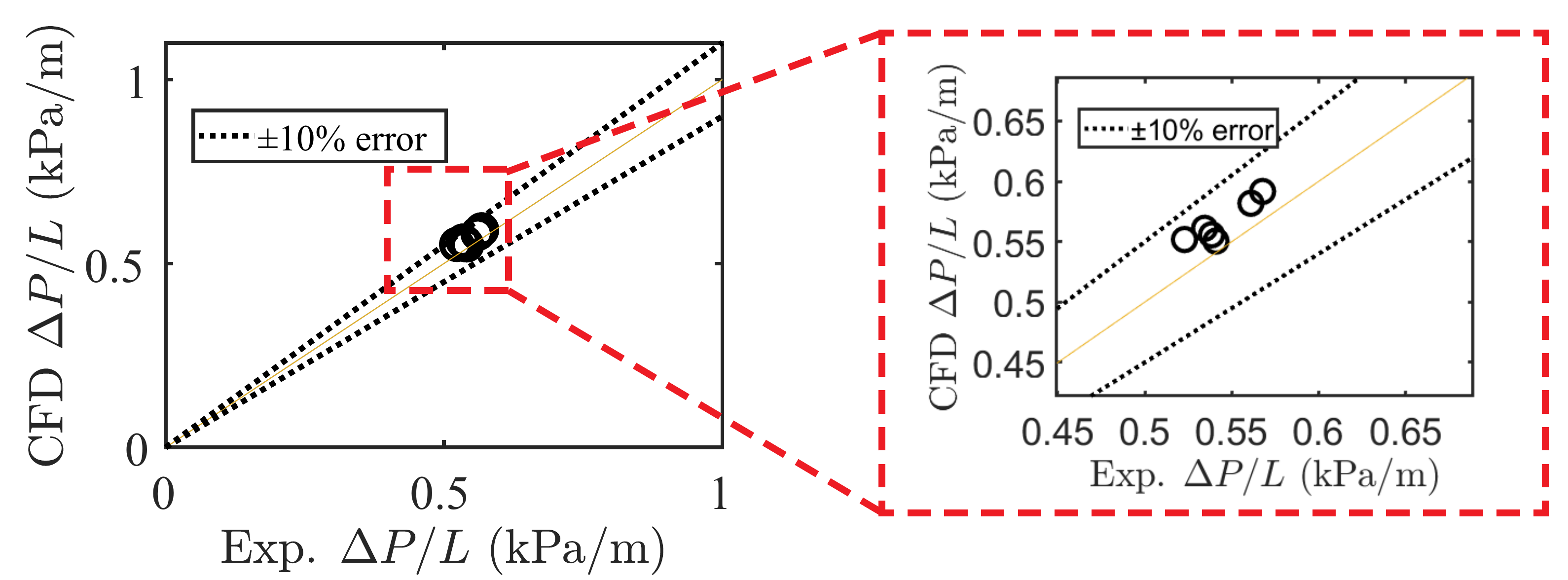}
	\caption{\label{fig:PressureCFD} Parity plot of pressure drop comparison with CFD predictions. The magnified view shows CFD vs. field data points with an error of less than 10 \%. Detailed data on all the fluid properties and flow conditions for model validation are provided in Table \ref{tab:Fielddata}.}
\end{figure}

Furthermore, the adequacy of the developed multiphase CFD model and the reliability of the prediction are also analyzed by the comparison of the CFD model predicted the pressure gradient and field data from the industrial pipeline, as portrayed in Fig.\ref{fig:PressureCFD}. The CFD model prediction demonstrated exceptional agreement with filed data with a maximum error lower than 10\%, indicating the CFD model efficacy. The magnified view of the six data sets agreement is also displayed in Fig.\ref{fig:PressureCFD}. In summary, the developed CFD model established the accuracy of forecast with filed data in terms of velocity field and pressure drop for different sets of data. As a result, the developed CFD model provides highly reliable predictions for industrial slurry systems with an acceptable error deviation.  
\section{Parametric study}
\subsection{Effect of bitumen droplet size}
\begin{figure}
	\centering
	\includegraphics[width=\textwidth]{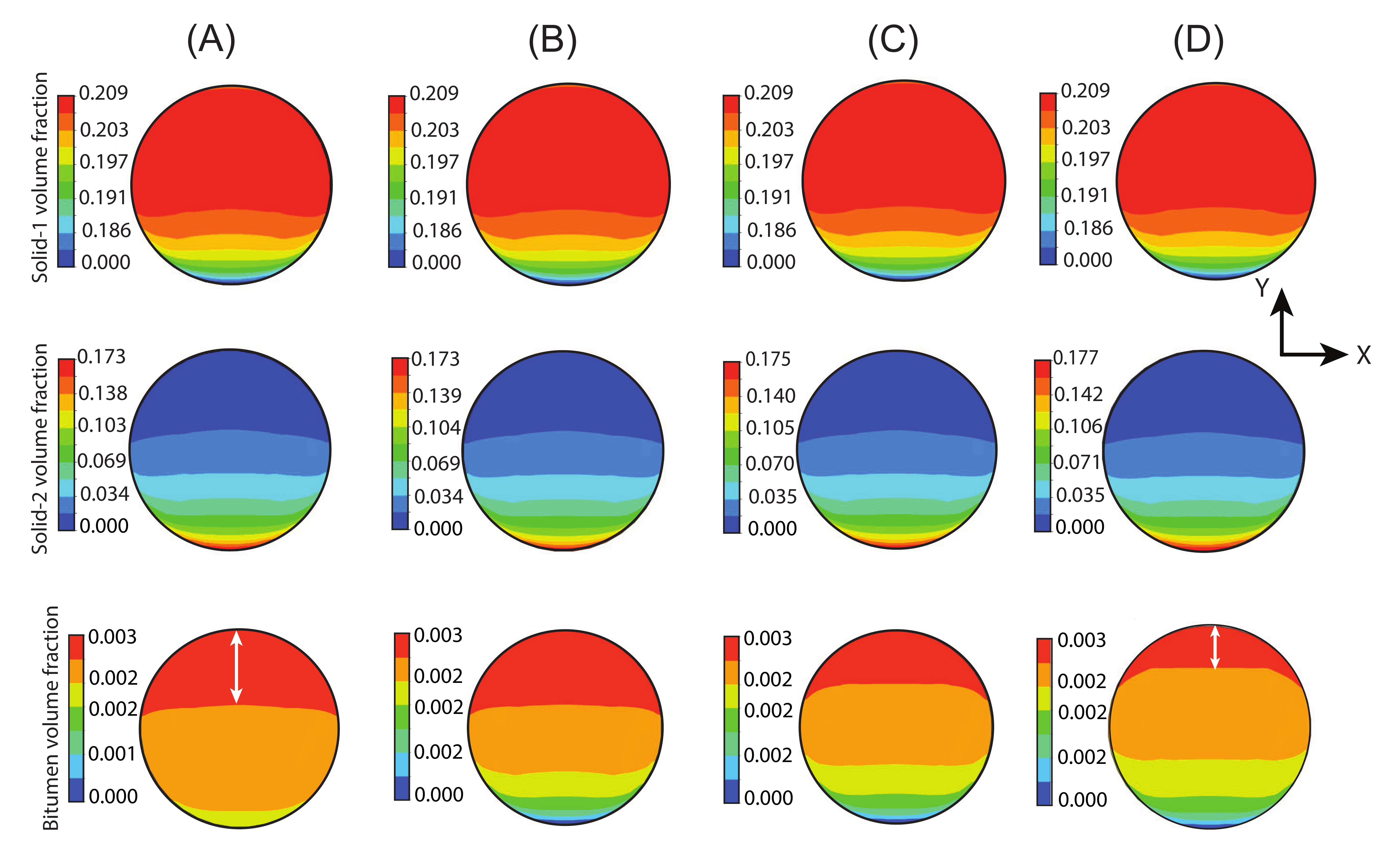}
	\caption{\label{fig:Bitumendrop1} The contour of solid\textendash 1, soild\textendash 2, and bitumen distribution crosses the pipe at Z = 100 m for different droplet sizes (A) 100 $\mu$m, (B) 200 $\mu$m, (C) 300 $\mu$m,  and (D) 400 $\mu$m.
	}
\end{figure}

In this section, the effect of bitumen droplet size in a tailing slurry system is systematically investigated. The influences of bitumen droplet size on solid particles distribution, velocity, solids, and bitumen profiles are studied at a fixed operating condition as reported in Table\ref{tab:Fielddata} for Case\textendash A. Fig.\ref{fig:Bitumendrop1} shows the small (i.e., soild\textendash 1), coarse (i.e., soild\textendash 2), and bitumen droplets distribution across the pipe at Z= 100 m.\\  

It is evident from Fig.\ref{fig:Bitumendrop1}A, that the solid concentration distribution is  completely different for both small and coarse particles. In the case of small particles, the solid fraction is mainly distributed from the top of the pipe to  the bottom part of the pipe. The coarse particles are accumulated at the bottom part of the pipe, where the small particle fraction is minimal. This is mainly due to the gravitational force acting on the large particles that leads to accumulate at the bottom part of the pipe. Gravitational forces result in particle distributions being symmetric in the horizontal direction but asymmetric vertically. \\

The bitumen distribution is analyzed at the same operating conditions, and it is found that bitumen droplets are mainly distributed from top to center of the pipe. This may be due to the fact that the small bitumen droplets accumulated at the top region in a highly turbulent flow similar to the small solid particles. In other words, the interaction force between the solid particles and bitumen droplets also changes bitumen droplet dynamics in a turbulent flow.\\

\begin{figure}
	\centering
	\includegraphics[width=\textwidth]{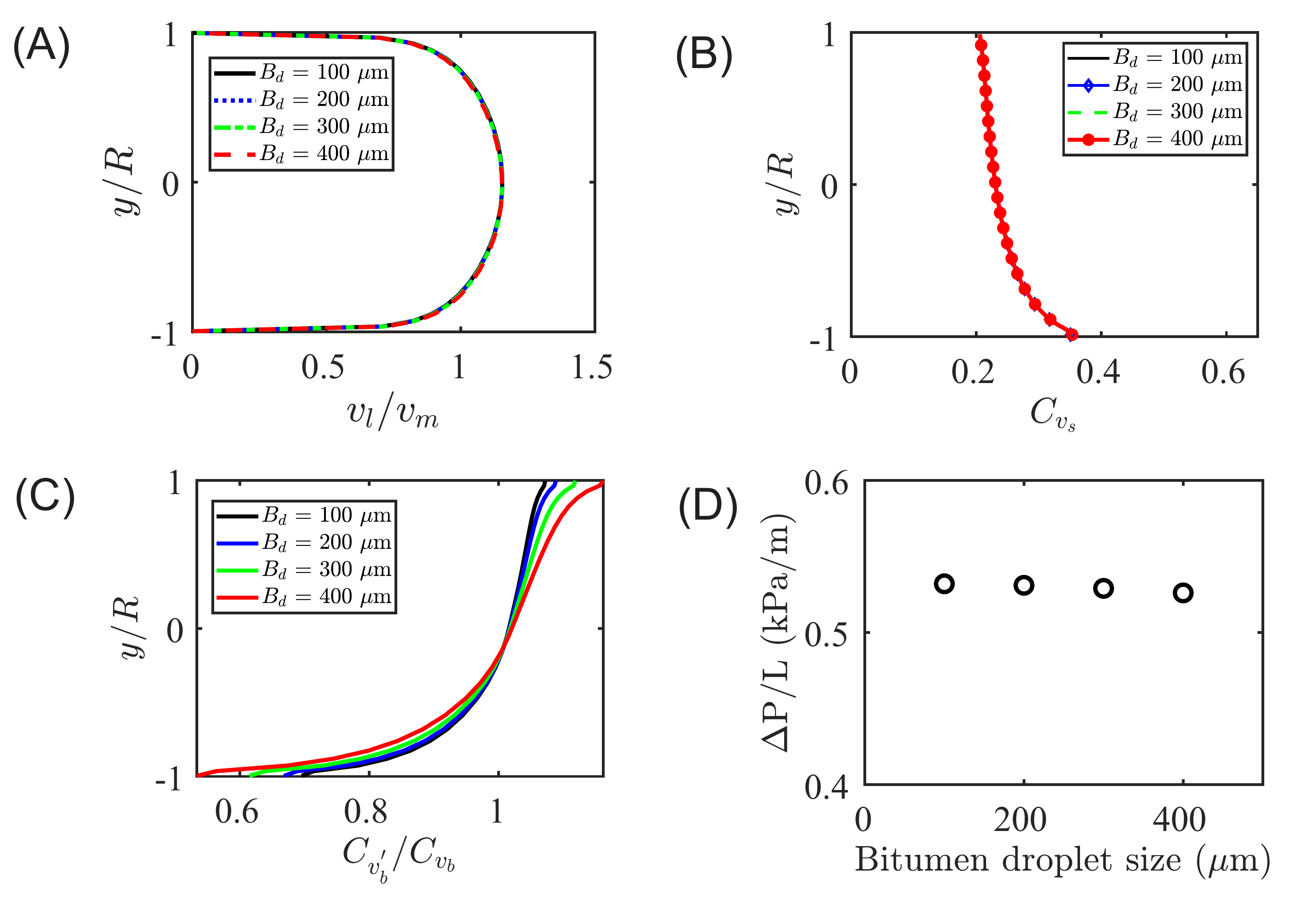}
	\caption{\label{fig:Bitumendrop2} Effect of bitumen droplet size on (A) velocity profiles, (B) chord\textendash average total solid concentration profiles, (C) bitumen distribution plots, and (D) pressure drop at fixed operation conditions.}
\end{figure}

The solid particle distribution with an increase in the bitumen droplet size from $B_d$ = 100 $\mu$m to 400 $\mu$m is almost identical to the qualitative observation, as shown in Fig.\ref{fig:Bitumendrop1}B\textendash D. For all the cases, the coarse particle concentration is relatively higher at the bottom part of the pipe. As a result, the dynamics of coarse particle motion are relatively lower at the bottom due to an increase in the particle\textendash particle and particle\textendash wall friction in the bottom region. The degree of asymmetry in the coarse particle distribution increases with increasing bitumen droplet size because the particle\textendash bitumen interactions are more significant, as shown in Fig.\ref{fig:Bitumendrop1}D. Therefore, the velocity in the bottom region is lower than the center and top part of the pipe compared to the low solid concentration region. However, with an increase in the bitumen droplet size, the bitumen concentration accumulated in a specific smaller region at the top part of the pipe. Moreover, this observation indicated that small size bitumen droplets distributed across the pipe are similar to small solid particles. However, this phenomenon is different when bitumen droplet size increases from 100 $\mu$m to 400 $\mu$m. This is mainly due to the amount of bitumen fraction in the domain being the same for all the conditions. Therefore, a larger bitumen accumulation is observed at the top region in contrast to a smaller bitumen droplet observation.\\

To determine the influence of bitumen droplet size on the flow characteristics such as velocity, chord\textendash average total solid concentration, bitumen distribution profiles are analyzed at Z = 100 m across the pipe as shown in Fig.\ref{fig:Bitumendrop2}A\textendash C. All the flow profiles are analyzed at the center line of the pipe from the bottom to the top. Fig.\ref{fig:Bitumendrop2}A indicates that the maximum velocity magnitude is observed at the center of the pipe, and velocity gradually decreases from the center of the pipe to the bottom and top with an increase in bitumen droplet size, the change in the carrier velocity profile is minimal. Numerical predicted carrier velocity profile trends also corroborated with the findings of slurry flow systems in pipelines from \citet{wang2013numerical} and \citet{bordet2018advanced}. This observation may be due to the low bitumen fraction for all the cases. The influence of bitumen droplet size on the initial bitumen concentration range may not be enough to alter the velocity profile characteristics. This suggests that the change in secondary phase droplet size has minimal effect on the velocity profile. \\

Fig.\ref{fig:Bitumendrop2}B represents the chord\textendash average total solid concentration profiles for the different bitumen droplet sizes. From the profiles, the effects from bitumen droplet size is negligible on the concentration profiles, as shown in Fig.\ref{fig:Bitumendrop2}B. The concentration gradient near the pipe bottom is higher compared to the pipe top region. A similar observation is reported in the bimodal particles in both experimental and numerical simulations \cite{kaushal2005effect,messa2020analysis,li2018hydrodynamic,li2018effect}. This may be due to solid particle accumulation at the bottom part of the pipe, and particle momentum is relatively low compared to the center of the pipe.\\


Fig.\ref{fig:Bitumendrop2}C depicts the bitumen distribution along a vertical line at a constant bitumen fraction in different bitumen droplet sizes. The bitumen profiles display that with an increase in the bitumen droplet size, the bitumen composition is relatively increased at the top part of the pipe. However, the bitumen fraction gradually decreases until the middle of the pipe. This implies that bitumen droplet size significantly influences the  distribution of bitumen. The bitumen distribution is almost identical for all the considered ranges of bitumen droplet size at the middle of the pipe. In the case of larger bitumen droplets, the bitumen fraction considerably lowered compared to smaller droplets  at the bottom of the pipe. This is because the coarse particles mainly accumulated at the bottom of the pipe, where the second phase interacts with coarse particles. This can result in smaller bitumen droplets being trapped between the particles in the bottom regime. This might be a consequence of the greater role played by particle\textendash particle and particle\textendash bitumen droplet interactions at the bottom part of the pipe.\\

Fig.\ref{fig:Bitumendrop2}D shows the frictional pressure drop for different bitumen droplet sizes. Results show that increasing bitumen droplet size results in a decrease in pressure drop. However, the change pressure drop is not prominent for the considered range of bitumen droplet size and operating conditions. In the case of smaller bitumen droplets. the interaction between secondary phases significantly contributes to a higher pressure drop due to the distribution of small particles and bitumen droplets from the top part of the pipe to the bottom part. In addition, the interaction between particle\textendash particle also contributes to increased pressure drop. 

\subsection{Effect of bitumen fraction}
\begin{figure}
	\centering
	\includegraphics[width=\textwidth]{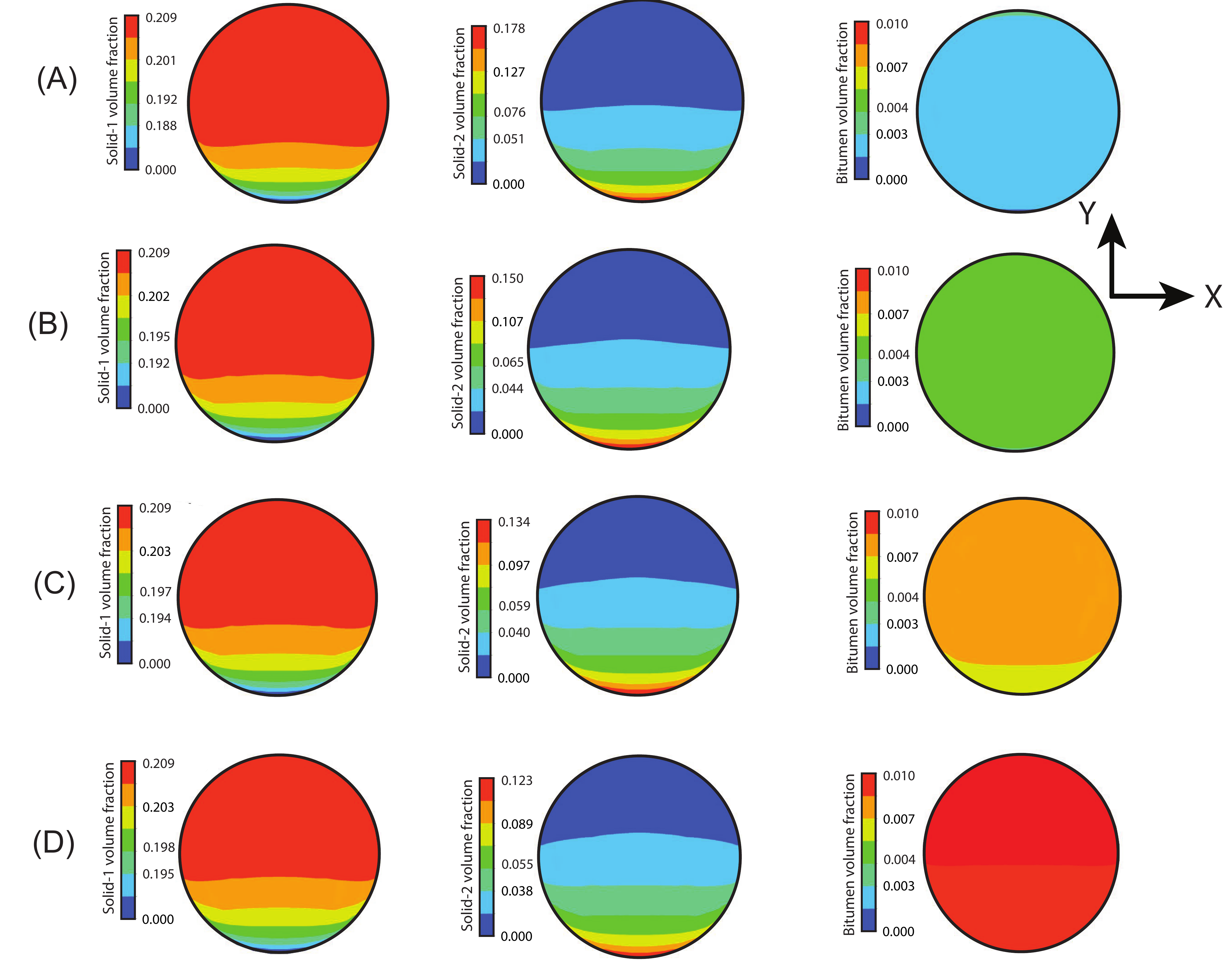}
	\caption{\label{fig:Bitumenfract1} The contour of solid and bitumen distributions.  Bitumen fraction:  (A) 0.0025, (B) 0.005, (C) 0.0075, and (D) 0.01.  Bitumen droplet size 400 $\mu$m.}
\end{figure}

To investigate the effect of bitumen fraction in tailings slurry systems, different bitumen compositions are considered from 0.0025 to 0.01, which is typically an industrial bitumen fraction range in tailings residuals. Fig.\ref{fig:Bitumenfract1} shows the solid particles and bitumen droplet distribution of contour plots for different bitumen fractions. The obtained numerical results reveal that for lower bitumen fraction conditions, small particles are distributed evenly from the top part of the pipe to the center of the pipe, as shown in Fig.\ref{fig:Bitumenfract1}A. The concentration of fine particles gradually increases in the bottom half of the pipe, and the asymmetry of the fine particle concentration profile along the vertical direction decreases. The gravitational force acts on coarse particles, causing most particles to accumulate at the bottom of the pipe. It is evident from top to center that the coarse fraction is minimal, while bitumen droplets and small particles are distributed throughout the pipe.\\

An increase in bitumen fraction showed negligible impact on the small particle distribution across the pipe at $Z = 100 m$ as shown in Fig.\ref{fig:Bitumenfract1}B\textendash D. The variation in coarse particle distributions is qualitatively similar to an increase in bitumen fraction, but the distribution of coarse particles height from the bottom to the top part of the pipe slightly decreased as expected. In other words, the solid concentration is nearly constant in the horizontal direction, and a noticeable variation in the vertical direction is observed as the bitumen fraction increases. The results indicate that coarse particle fraction distribution increases when bitumen fraction increases. This may be due to an increase in the viscosity of the slurry system compared to the lower bitumen composition. However, the change in coarse solid (i.e., solid\textendash 2) fraction distribution clearly evident from Fig.\ref{fig:Bitumenfract1}A and D. Notably, the concentration profile of fine particles displays a distribution trend opposite to that of coarse particles with an increase in bitumen droplet size.\\      

Fig.\ref{fig:Bitumenfract1}A\textendash D  contours also show the bitumen fraction distribution across the pipe at $Z = 100 m$ by altering the initial bitumen fraction up to 0.01.  The bitumen distribution is significantly higher above the pipe center and gradually decreases from the center to the bottom part of the pipe. With an increase in bitumen fraction, the results demonstrate the presence of bitumen droplets at the bottom of the pipe. The maximum distribution observed for higher bitumen fraction composition may be due to an increased particle\textendash bitumen interaction in the domain and also leads increase in coarse particle distribution.\\

\begin{figure}
	\centering
	\includegraphics[width=\textwidth]{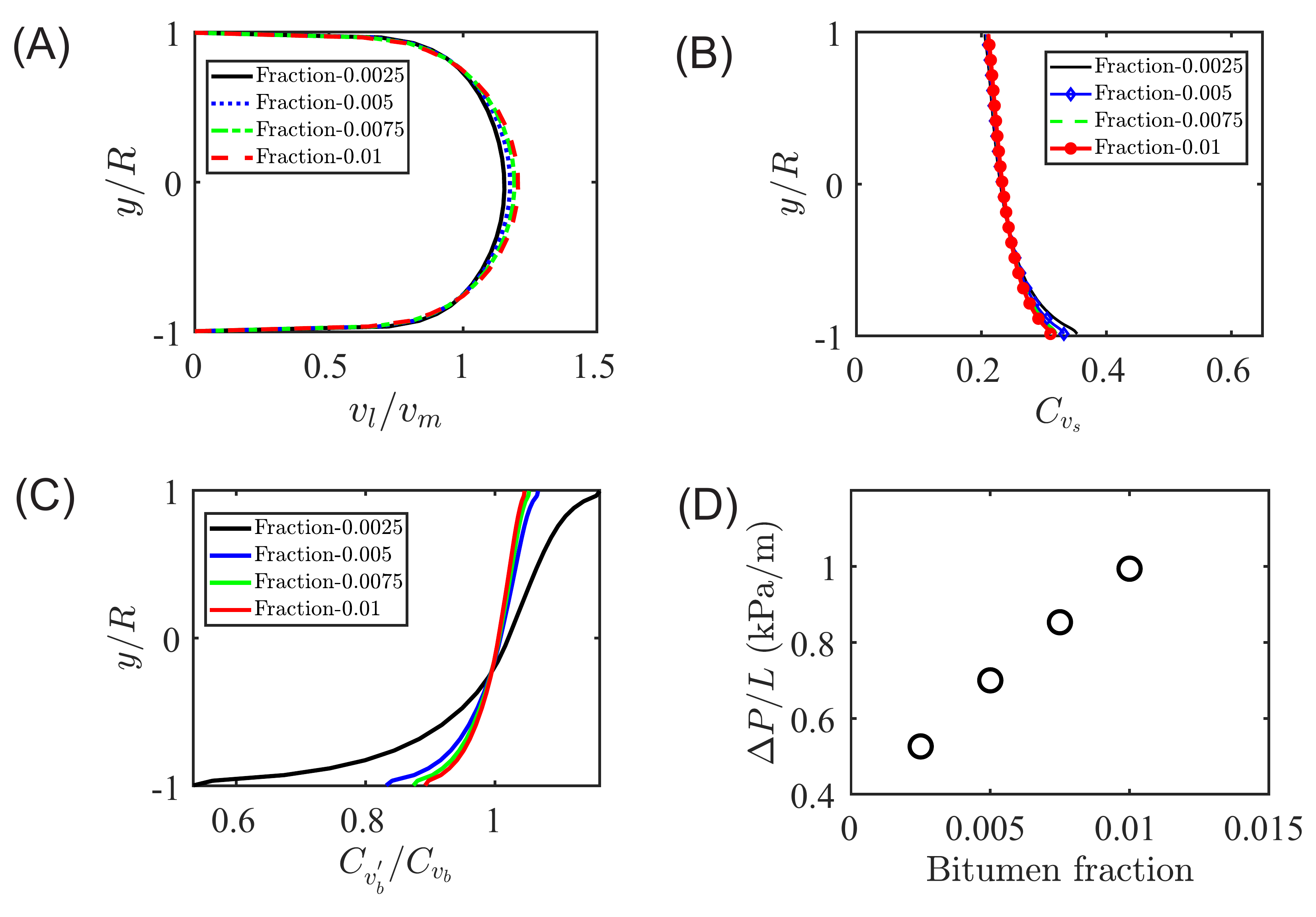}
	\caption{\label{fig:Bitumenfract2} Effect of bitumen fraction on (A) velocity profiles, (B) chord\textendash average total solid concentration profiles, (C) bitumen distribution plots, and (D) pressure drop at fixed operation conditions. Bitumen droplet size 400 $\mu$m.}
\end{figure}

Fig.\ref{fig:Bitumenfract2} illustrates the quantitative analysis of flow profiles and pressure drop for different bitumen fraction conditions by keeping all other conditions similar. It can be seen that with an increased bitumen fraction carrier velocity profile slightly changed due to the change in viscosity of the slurry system and particle\textendash bitumen interactions. The presence of a higher bitumen fraction corresponds to more bitumen droplets in the slurry system due to higher turbulence characteristics. As a result, the velocity magnitude is relatively higher in the center of the pipe for a higher bitumen fraction compared to other conditions. However, the velocity profiles reveal that top and bottom parts of the pipe, the velocity distribution is more symmetrical, as shown in Fig.\ref{fig:Bitumenfract2}A.\\

In Fig.\ref{fig:Bitumenfract2}B, the chord\textendash averaged solid fraction is presented to evaluate the effect of the bitumen fraction. The results indicated that the solid fraction profile gradually decreased from the top part of the pipe to the bottom part. As stated earlier, the gravitational force acting on coarse particles leads to accumulation at the bottom of the pipe along with small particles due to particle\textendash particle interactions. It is worth noting that the solid fraction distribution slightly decreased near the wall with an increase in bitumen fraction. This is in accord with the solids fraction distribution information across the pipe at $Z = 100 m$ as depicted in Fig.\ref{fig:Bitumenfract1}. The results show that secondary phase interactions strongly influence the solid particle distribution in highly turbulent and viscous flows.\\

Fig.\ref{fig:Bitumenfract2}C provides an analysis of bitumen distribution for different ranges of bitumen fractions. It is found that with an increase in bitumen fraction, the distribution is more homogeneous in the system. However, for smaller bitumen fraction cases, a noticeable difference is observed in the bitumen fraction profiles from the top to the bottom of the pipe. In the case of lower fractions, more bitumen is accumulated at the top part of the pipe due to less interaction between solid particles and bitumen droplets. But the bitumen fraction distribution is significantly lowered at the bottom of the pipe. On the other hand, for higher bitumen fraction cases, the bitumen distribution is slightly higher at the bottom of the pipe compared to lower bitumen fraction cases.\\

In addition, the pressure drop is also analyzed for all the conditions. Fig.\ref{fig:Bitumenfract2}D demonstrates the pressure drop for different bitumen fractions by keeping other conditions, and bitumen droplet sizes are constant. It is profound that the increase in bitumen fraction significantly contributes to increase in pressure drop of the system. The particle\textendash particle interaction and particle\textendash bitumen interactions tend to concentrate more intensively from the top to bottom part of the pipe. Another possible explanation for this change in pressure drop is that the viscosity of the slurry system can also change with bitumen fraction. The increase in pressure drop resulted in an increase in specific power consumption and eventually more pumping cost in the industrial scale slurry transport in pipelines.  

\subsection{Effect of bubble size}
\begin{figure}
	\centering
	\includegraphics[width=\textwidth]{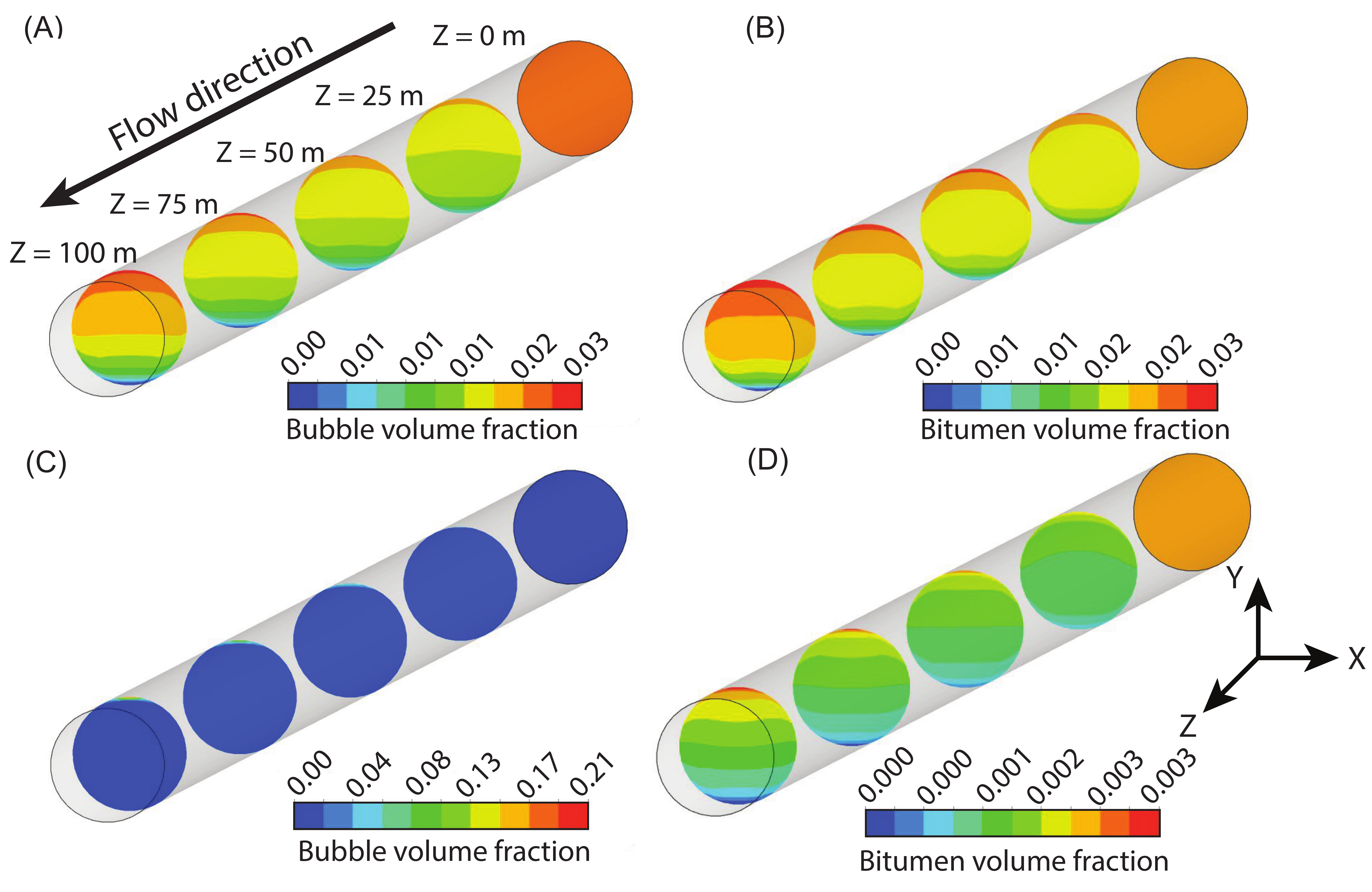}
	\caption{\label{fig:Bubblesize1} (A) Bubble distribution, and (B) bitumen distribution for 5 $\mu$m bubble size. (C) Bubble distribution, and (D) bitumen distribution for 1000 $\mu$m bubble size.}
\end{figure}

This section discusses the influence of gas bubble size in the slurry system for fixed conditions. Different sizes of gas bubbles are considered ranging from 5 $\mu$m to 1000 $\mu$m based on the available experimental data. To understand the effect of bubble size on flow characteristics, bubble fraction is considered similar to bitumen fraction, as mentioned in Table.\ref{tab:Fielddata} case\textendash A. Fig.\ref{fig:Bubblesize1} shows the distribution of bubbles and bitumen droplets along the Z\textendash length of the channel for two cases. All the analyses are performed after reaching the stable flow, and the flow time is at $200$ $ s$. The simulation results demonstrate that the gas bubble gradually moved to the pipe's top part along the pipe's length, as shown in Fig.\ref{fig:Bubblesize1}A. However, it is evident from Fig.\ref{fig:Bubblesize1}B that the bitumen droplets are uniformly distributed at the inlet, and the bitumen droplets segregation gradually changes along with the length of the channel. Due to the smaller contact area between the tiny bubble and the particle, tiny bubbles attach to particles/droplets more easily and faster than larger bubbles.\cite{zhou2020role}\\ 

The evolution of bitumen droplets is relatively high in the middle of the pipe, where Z = 75 m, and the bitumen is distributed from the top to the bottom part of the pipe. Near the outlet, most of the bitumen is accumulated at the top in a specific region due to bubble\textendash bitumen interactions and particle\textendash bitumen interactions. Importantly, a similar phenomenon is observed when the bubble size is 1000 $\mu$m, but the gas bubbles are accumulated at the top part of the pipe in one specific region. Fig.\ref{fig:Bubblesize1}C displays the gradual change in bubble distribution along the length of the channel. The results clearly show that minimal gas bubble distribution is observed from the center to the bottom part of the pipe, significantly different from the smaller bubble size case distributions. Recently, experimental work of \citet{rosas2018measurements}  reported that the smaller dispersed bubbles remain stable in the complex slurry system since the buoyance forces are not sufficient to overcome the turbulent forces when the bubble diameter is sufficiently small. The numerical findings also corroborated with the \citet{rosas2018measurements} work for different bubble sizes as shown in Fig.\ref{fig:Bubblesize1}A and c. In other words, the gravitational force that acts on the small bubbles is negligible compared to large\textendash size bubbles. In the case of large\textendash size bubbles, the buoyance forces are more prominent to overcome the turbulence forces. The bitumen distribution trend for a larger bubble is very similar to the 5 $\mu$m bubble case, as displayed in Fig.\ref{fig:Bubblesize1}D.\\          
\begin{figure}
	\centering
	\includegraphics[width=\textwidth]{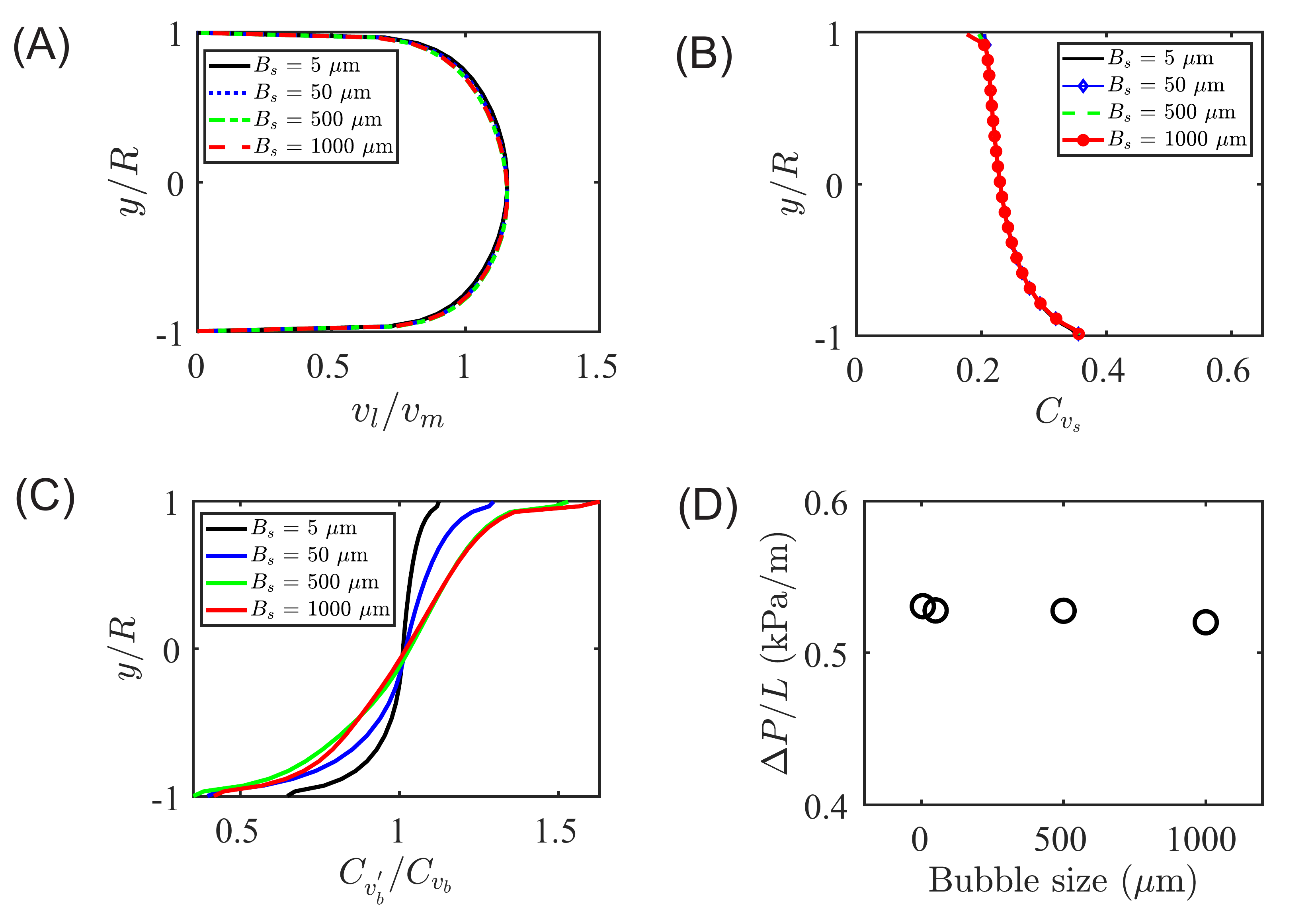}
	\caption{\label{fig:Bubblesize2} Effect of bubble size (A) velocity profiles, (B) chord\textendash average total solid concentration profiles, (C) bitumen distribution plots, and (D) pressure drop at fixed operation conditions. $B_{s}$ denotes the bubble size.}
\end{figure}

To further understand the .influence of gas bubbles, flow profiles, and pressure drop are quantified, as shown in Fig.\ref{fig:Bubblesize2}. For all the cases, carrier velocity profiles showed a negligible change as shown in Fig.\ref{fig:Bubblesize2}A.
It can be seen from Fig.\ref{fig:Bubblesize2}B that for all the cases, chord\textendash average total solid concentration profiles are similar. This suggests that bubble size has a negligible effect on solid concentration profiles for the considered range of bubble fractions. It is imperative to mention that concentration profiles are similar, but small and coarse particle distributions may differ for similar conditions.\\

Fig.\ref{fig:Bubblesize2}C depicts the bitumen distribution profiles at the centerline of the plane for different bubble sizes. It is qualitatively perceptible that the amount of bitumen is relatively higher for larger bubbles and  gradually to a minimum value. It is worth noting that the bitumen distribution profiles are similar for both 500 $\mu$m and 1000 $\mu$m bubble sizes. However, it is important to notice that for smaller bubble sizes 5 $\mu$m and 50 $\mu$m case, the amount of bitumen is lower at the pipe's top part compared to larger bubbles. \citet{booth1954flotation} also reported a similar observation on the role of these tiny bubbles in accelerating particle\textendash bubble/droplet attachment and improving recovery.  Therefore, the bubble size also plays a significant role in the bitumen distribution in the pipe for efficient recovery. Notably, this finding indicated a strong synergy between bubbles and particles/ bitumen droplets. On the other hand, the pressure drop decreased with an increase in bubble size, as shown in Fig.\ref{fig:Bubblesize2}D.      

\subsection{Effect of bubble fraction}
\begin{figure}
	\centering
	\includegraphics[width=\textwidth]{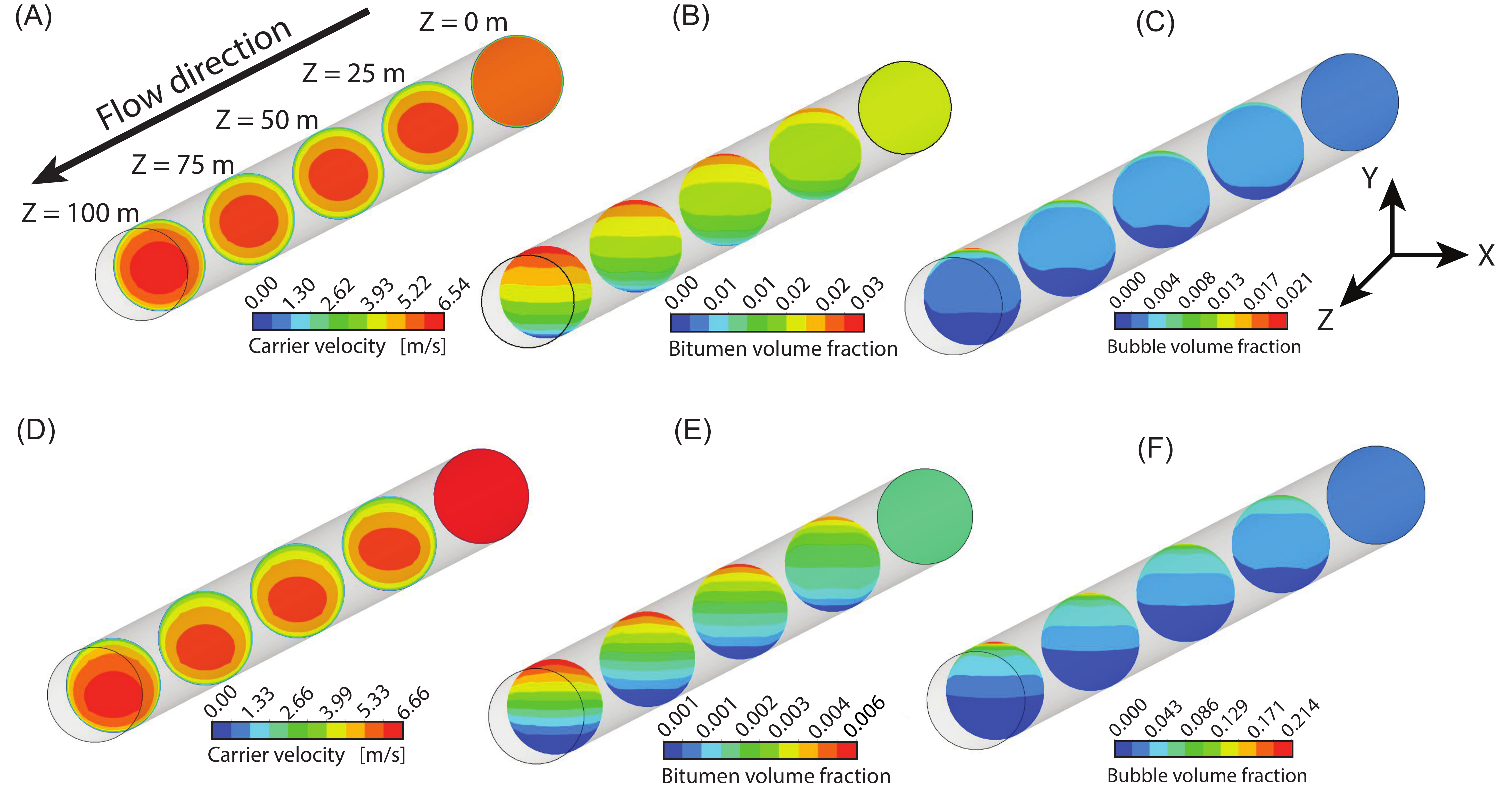}
	\caption{\label{fig:Bubblefrac1} (A) Carrier velocity distribution, (B) bitumen distribution, and (C) bubble distribution for bubble fraction 0.0025. (D) Carrier velocity distribution, (E) bitumen distribution, and (F) bubble distribution for bubble fraction 0.03. Bubble size 500 $\mu$m.}
\end{figure}

This section demonstrates the effect of bubble formation on the pipeline's bitumen and gas bubble distribution. Numerical simulations are conducted at bubble fractions ranging from 0.0025 to 0.03 by considering the fixed bubble size of 500 $\mu$n and bitumen droplet size of 400 $\mu$m. Fig.\ref{fig:Bubblefrac1} shows the influence of bubble fraction on the carrier velocity, bitumen, and bubble distribution along the flow directions. It can be seen from Fig.\ref{fig:Bubblefrac1}A that the carrier velocity distribution gradually changed toward the outlet. The carrier velocity profile reaches stable flow and developed flow profile conditions after reaching critical length Z = 75 m.\\

The maximum velocity distribution is at the center of the pipe after reaching the critical length. The particle\textendash particle collision and turbulent dispersion of particle clusters near the top wall result in a steep velocity gradient from the wall to the center region. In line with previous observations, the maximum amount of bitumen  is accumulated on the top part of the pipe, and the distribution gradually decreases from the top to the bottom part of the pipe as displayed in Fig.\ref{fig:Bubblefrac1}B. Gas bubbles also accumulated at the upper part of the pipe in a specific region, as shown in Fig.\ref{fig:Bubblefrac1}C. In contrast, with an increased bubble fraction, the carrier velocity distribution is shifted to the bottom side of the pipe, and the upper portion velocity magnitude range is relatively lower. It could be attributed to an increase in bubble fraction since bitumen, bubbles, and fine particles are accumulated in the upper region, as demonstrated in Fig.\ref{fig:Bubblefrac1}D.\\

\begin{figure}
	\centering
	\includegraphics[width=0.85\textwidth]{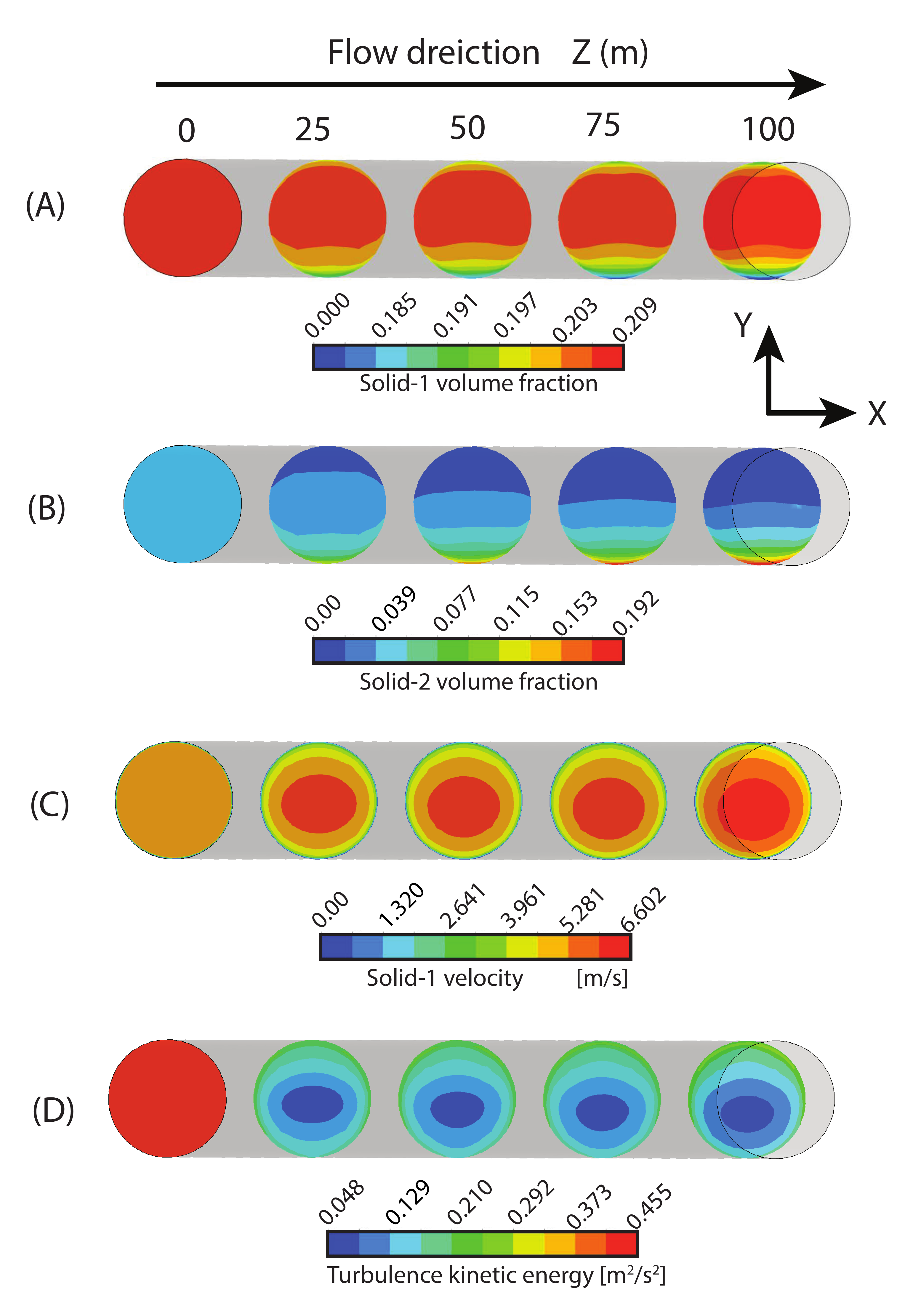}
	\caption{\label{fig:Bubblefrac3} (A) solid\textendash 1  fraction, (B) solid\textendash 2  fraction, (C) solid\textendash 1 velocity, and (D) turbulence kinetic energy distribution along the the Z\textendash direction for bubble fraction 0.01 and bubble size 500 $\mu$m.}
\end{figure}

Similar to the results from smaller bubble fractions, with an increase in bubble fraction at fixed conditions, the bitumen and bubbles distribution are identical with a change maximum fraction range at the upper part of the pipe. The higher number of bubbles in the domain is due to a change in bubble fraction. Consequently, the higher number of bubbles critically impacts the bitumen accumulation at the top part of the pipe, as depicted in Fig.\ref{fig:Bubblefrac1}E. It is evident from Fig.\ref{fig:Bubblefrac1}F that the bubble distribution is significantly different from the smaller bubble fraction case. With an increased bubble fraction, most bubbles are accumulated at the top and distributed to the center of the pipe. 
Fig.\ref{fig:Bubblefrac3}A\textendash D also shows the three\textendash dimensional view of particle and bitumen distribution across the different cross\textendash sections from the inlet to the outlet. It can be observed from Fig.\ref{fig:Bubblefrac3}A that small particle accumulation reached a stable flow pattern, and also along the length, coarse particle accumulation increased until Z = 100 m (Fig.\ref{fig:Bubblefrac3}B). The corresponding small particle velocity  developed a fully developed flow condition near the outlet, as shown in Fig.\ref{fig:Bubblefrac3}C, and turbulence kinetic energy also slightly increased along the flow direction (Fig.\ref{fig:Bubblefrac3}D).\\  
\begin{figure}
	\centering
	\includegraphics[width=\textwidth]{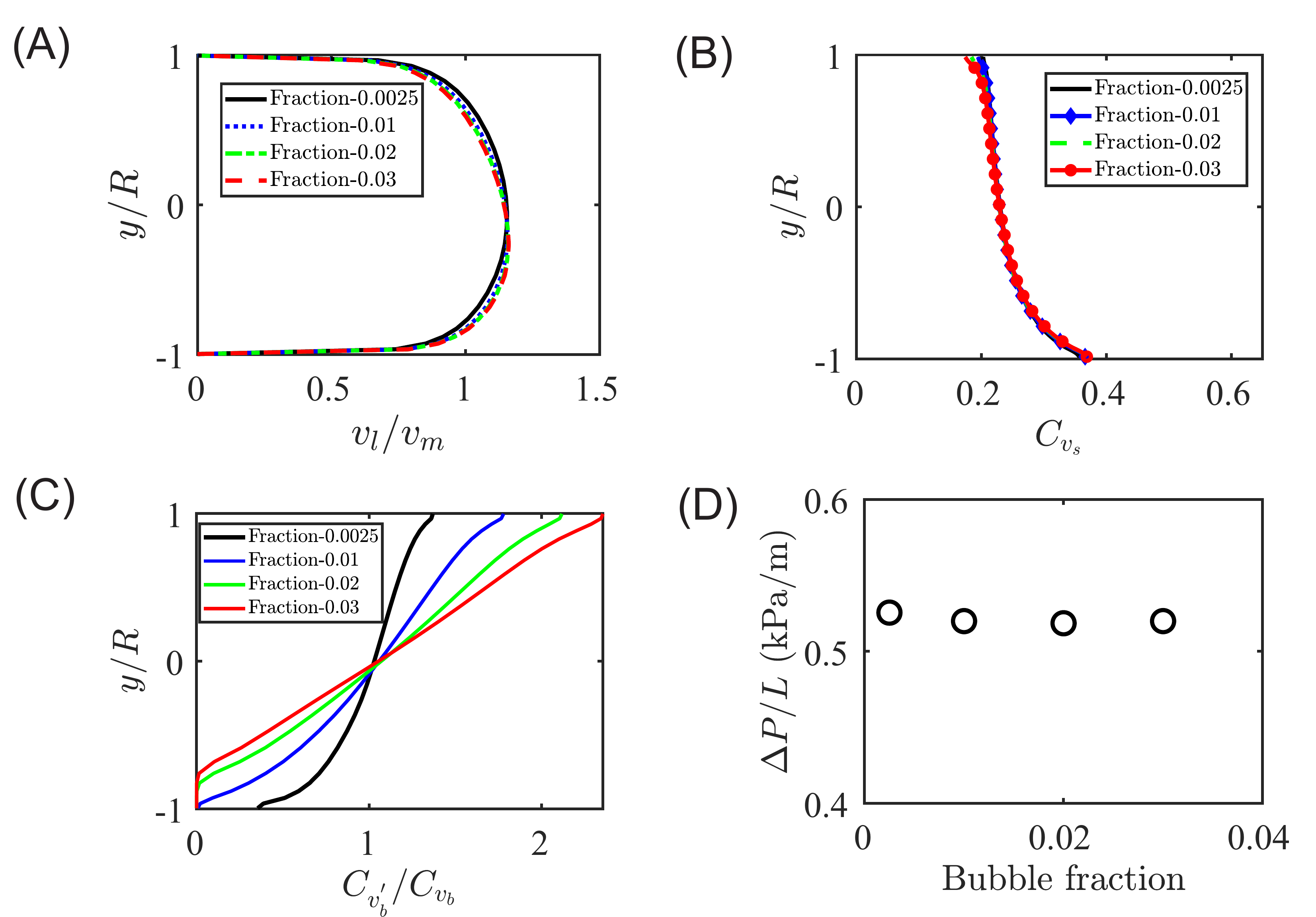}
	\caption{\label{fig:Bubblefrac2} Effect of bubble fraction on (A) velocity profiles, (B) chord\textendash average total solid concentration profiles, (C) bitumen distribution plots, and (D) pressure drop at fixed operation conditions.}
\end{figure}

Fig.\ref{fig:Bubblefrac2}A demonstrates the effect of bubble fraction on carrier velocity profile across the vertical centerline for different bubble fraction conditions. A change in velocity profiles is observed with an increase in bubble fraction from 0.0025 to 0.03. At a higher bubble fraction, the carrier velocity profile is shifted slightly downward for the channel, and the velocity magnitude is relatively higher at the bottom part of the pipe. This is mainly due to the number of gas bubbles and bitumen droplets being greater at the top part of the pipe, which leads to a higher velocity magnitude at the bottom part of the pipe. For different bubble fraction conditions, Fig.\ref{fig:Bubblefrac2}B shows similar trends in solid volume fraction profiles. With an increase in bubble fraction, the top part of the pipe solids concentration decreased compared to conditions with a lower bubble fraction. Specifically, this is due to a higher concentration of bubbles and bitumen droplets.\\

\begin{table}
	\caption{Bitumen recovery summary.}
	\vspace{0.05cm}
	\label{tab:Recovery}
	\centering
	\renewcommand{\arraystretch}{1.4}
	\begin{tabular}{|c|c|ccc|}
		\hline
		\multirow{2}{*}{Study}                         & \multirow{2}{*}{Parameter range} & \multicolumn{3}{c|}{Bitumen recovery percentage}                           \\ \cline{3-5} 
		&                                  & \multicolumn{1}{c|}{Top 25 \%} & \multicolumn{1}{c|}{Top 50 \%} & Top 75\% \\ \hline
		\multirow{4}{*}{Bitumen droplet size ($\mu$m)} & 100                              & \multicolumn{1}{c|}{27.70}     & \multicolumn{1}{c|}{52.68}     & 78.15    \\ \cline{2-5} 
		& 200                              & \multicolumn{1}{c|}{27.91}     & \multicolumn{1}{c|}{52.93}     & 78.39    \\ \cline{2-5} 
		& 300                              & \multicolumn{1}{c|}{28.34}     & \multicolumn{1}{c|}{53.49}     & 78.90    \\ \cline{2-5} 
		& 400                              & \multicolumn{1}{c|}{28.98}     & \multicolumn{1}{c|}{54.32}     & 79.66    \\ \hline
		\multirow{4}{*}{Bitumen fraction}              & 0.0025                           & \multicolumn{1}{c|}{28.98}     & \multicolumn{1}{c|}{54.32}     & 79.66    \\ \cline{2-5} 
		& 0.005                            & \multicolumn{1}{c|}{27.31}     & \multicolumn{1}{c|}{51.91}     & 77.13    \\ \cline{2-5} 
		& 0.0075                           & \multicolumn{1}{c|}{27.06}     & \multicolumn{1}{c|}{51.54}     & 76.76    \\ \cline{2-5} 
		& 0.01                             & \multicolumn{1}{c|}{26.93}     & \multicolumn{1}{c|}{51.35}     & 76.57    \\ \hline
		\multirow{4}{*}{Bubble size ($\mu$m)}                   & 5                                & \multicolumn{1}{c|}{27.97}     & \multicolumn{1}{c|}{52.75}     & 78.24    \\ \cline{2-5} 
		& 50                               & \multicolumn{1}{c|}{30.36}     & \multicolumn{1}{c|}{55.82}     & 81.04    \\ \cline{2-5} 
		& 500                              & \multicolumn{1}{c|}{32.78}     & \multicolumn{1}{c|}{59.03}     & 83.22    \\ \cline{2-5} 
		& 1000                             & \multicolumn{1}{c|}{32.80}     & \multicolumn{1}{c|}{58.70}     & 82.43    \\ \hline
		\multirow{4}{*}{Bubble fraction}               & 0.0025                           & \multicolumn{1}{c|}{32.78}     & \multicolumn{1}{c|}{59.03}     & 83.22    \\ \cline{2-5} 
		& 0.01                             & \multicolumn{1}{c|}{39.72}     & \multicolumn{1}{c|}{68.52}     & 90.64    \\ \cline{2-5} 
		& 0.02                             & \multicolumn{1}{c|}{45.65}     & \multicolumn{1}{c|}{76.22}     & 95.78    \\ \cline{2-5} 
		& 0.03                             & \multicolumn{1}{c|}{49.32}     & \multicolumn{1}{c|}{80.51}     & 97.68    \\ \hline
	\end{tabular}
\end{table}

Subsequently, the frequency of random collision increases during the movement of bubbles and droplets along with small particles to the top part of the pipe. The accumulation of particles with droplets and bubbles enhances the interactions and promotes the distribution of bitumen at the top part of the pipe. As displayed in Fig.\ref{fig:Bubblefrac2}C, bitumen profiles appear significantly altered with higher bubble fractions. Higher bubble fractions result in maximum bitumen accumulation and a linear decrease in bitumen profile from top to bottom. At higher bubble fractions, the number of gas bubbles is comparatively higher, and the frequency of bubble\textendash bitumen interactions is also accelerated, which causes bitumen to accumulate at the top of the pipe. The pressure drop with a higher bubble fraction, as shown in Fig.\ref{fig:Bubblefrac2}D.\\

To understand the influence of gas bubble size and fraction on bitumen recovery, for all the systematic numerical investigations, bitumen recovery is estimated at Z = 100 m. A vertical center line is considered on the cross\textendash section plane, and bitumen fraction recovery is estimated at different levels from top to bottom of the pipe specified in Table \ref{tab:Recovery}. The bitumen recovery is found to increase with increased bitumen droplet size, and from the top to the middle of the pipe, the maximum bitumen recovery is approximately 55 \%. However, the bitumen recovery greatly improved to approximately 80 \%. The optimum bitumen droplet size of 400 $\mu$m showed maximum bitumen recovery for the considered operating conditions. The higher bitumen fraction negatively impacted bitumen recovery from 54.32 \% to 51.35\% for the top 50\% cross\textendash section. Thus, the optimum bitumen fraction of 0.0025 showed maximum recovery under the fixed operating conditions.\\

Furthermore, bitumen recovery is also assessed with various gas bubble sizes, and  the optimum bubble size is determined to be  500 $\mu$m. The maximum bitumen recovery of 59\% could be observed for  500 $\mu$m, and a subsequent increase in bubble size bitumen recovery marginally decreased. The bubble fraction showed a significant effect on bitumen recovery with the bubble size of 500 $\mu$m. With an increase in bubble fraction, the bitumen recovery increased from a lower value of 59 \% to 80 \%. Therefore, CFD results demonstrate the bubble fraction plays a critical role in the bitumen recovery for the considered conditions. 
\cleardoublepage
\section{Conclusions}
In this work, we employ a three\textendash dimensional, transient Eulerian  CFD model to study the flow behavior of complex multiphase slurry systems. Four\textendash phases of non\textendash Newtonian tailings slurry flow  with bitumen droplets and bubbles in an industrial pipeline are modeled.  The CFD model is validated with industrial field data for 6 sets in terms of velocity profile and pressure drop with a maximum error of  6\% and $<$10\%. A detailed sensitivity analysis is demonstrated on the selection of carrier\textendash solid and solid\textendash bitumen drag models.  The combination of small and large particle sizes (i.e., 75 \& 700 $\mu$m) and bitumen droplet size (i.e., 400 $\mu$m) provided good agreement with field data in velocity profile and pressure drop.\\

Our numerical findings reveal that the bitumen droplet size plays a significant role in bitumen distribution, and larger droplets accumulate at the top part of the pipe. Bitumen droplet size strongly influences bitumen distribution profiles. With an increase in bitumen fraction, solid concentration profiles slightly shifted, and pressure drop increased. This study revealed that with an increase in bitumen fraction, pressure drop increased, and bitumen distribution profiles also showed significant differences due to a change in slurry composition. However, the coarse particle distribution also changed with an increase in bitumen fraction from 0.0025 to 0.01. The results indicates that with an increase in bubble size, bitumen distribution effectively improved, and the optimum bubble size is noted as 500 $\mu$m. Higher bubble fractions showed a strong influence on velocity and concentration profiles. The optimum conditions for higher bitumen recovery are revealed by CFD results in the pipeline.\\

The developed CFD model provides a powerful tool for understanding the complex multiphase flow behaviors during highly turbulent and viscous slurry transport. Therefore, this work contributes toward accurate predictions that may guide the process design of an industrial\textendash scale slurry transport. The outcomes of these studies are likely to guide conditions for bitumen recovery from tailings slurries.\\

\section*{Declaration of competing interest}
The authors declare that they have no known competing financial interests or personal relationships that could have appeared to influence the work reported in this paper.

\section*{Acknowledgement}

The authors acknowledge the funding support from the
Institute for Oil Sands Innovation (IOSI) (Project IOSI 2019\textendash 04 (TA)) and from the Natural Science and Engineering
Research Council of Canada (NSERC)\textendash Alliance. This research
was undertaken, in part, thanks to funding from the Canada Research Chairs Program. We also thank Compute Canada (www.computecanada.ca) for continued support through
extensive access to the Compute Canada HPC Cedar and
Graham clusters.

\section*{Credit author statement}

\textbf{Somasekhara Goud Sontti:} Methodology, Planned and performed the simulations, Model validation, Formal analysis, Visualizations, Writing\textendash original draft, Writing\textendash review \& editing. \textbf{Mohsen Sadeghi:} Model validation, Writing\textendash review \& editing. \textbf{Kaiyu Zhou:} Field data analysis. \textbf{Enzu Zheng:} Writing\textendash review \& editing.  \textbf{Xuehua Zhang:} Conceptualization, Methodology, Project administration, Writing\textendash review \& editing, Resources, Supervision.

\section*{Data availability}
The data that support the findings of this study are available from the corresponding author upon reasonable request.
\cleardoublepage
\section*{Nomenclature}

\begin{longtable}{l p{12cm}}
	$D$ &  pipe diameter (L) \\
	$R$ & pipe radius (L)\\
	$d_p$ & particle diameter (L)\\
	$C_{v}$ & chord\textendash averaged concentration (--)\\
	$g$ & gravitational acceleration (L T$^{-2}$)\\
	$g_0$ & radial distribution function (--)\\
	$p$ & locally\textendash averaged pressure (M L$^{-1}$ T$^{-2}$)\\
	$t$ & time (T) \\
	$v$ & velocity (L T$^{-1}$)\\
	$V$ & velocity (L T$^{-1}$)\\
	$f_\mathrm{drag}$ & drag function (--)\\
	
	$C_\mathrm{fr}$ & friction coefficient between solid phases (--)\\
	$x$ & horizontal coordinate (L)\\
	$y$ & vertical coordinate (L)\\
	$z$ & axial coordinate (L)\\
	$e$ & restitution coefficient (--)\\
	$I_{2D}$ & second invariant of the deviatoric stress tensor (--)\\
	$\norm{\vec{v}_s^{\,\prime}}$ & fluctuating solids velocity (L T$^{-1}$)\\
	$i$ & hydraulic gradient (--)\\
	$K_{ls}$ & momentum exchange coefficient between fluid\\
	$\Delta P$ & area\textendash averaged gauge pressure (M L$^{-1}$ T$^{-2}$))\\
	$k$& turbulent kinetic energy (L$^{2}$ T$^{-2}$)) \\ \\
	
	\textit{Greek symbol}\\ 
	
	$\alpha$ & locally\textendash averaged volume fraction (--)\\
	$\mu$ & dynamic viscosity (M L$^{-1}$ T$^{-1}$)\\
	$\rho$ & density (M L$^{-3}$)\\
	$\phi_{ls}$ & the energy exchange between the fluid and the solid phases (E) \\
	$\gamma_{\Theta_{s}}$ & collisional dissipation of energy (E)\\
	$\tau$ & shear stress (M L$^{-1}$ T$^{-2}$)\\
	$\dot{\gamma}$ & shear strain rate (T$^{-1}$)\\
	$\alpha_{s,\mathrm{max}}$ & maximum packing limit (--)\\
	$\Theta$  & granular temperature (L$^{-2}$ T$^{-2}$)\\
	$\varphi$ & angle of internal friction (--)\\
	$\eta_{t}$ & turbulent diffusivity (--)\\
	$\eta$ & apparent viscosity (M L$^{-1}$ T$^{-1}$)\\ \\
	
	\textit{Subscripts}\\
	
	$l$ & liquid\\
	$s$ & solid\\
	$ss$ & solid particles \\
	$p$ & $p^{th}$ solid phase\\
	$q$  & $q^{th}$ solid phase\\
	col & collisional part of viscosity\\
	kin & kinetic part of viscosity\\
	fr & frictional part of viscosity  \\ 
	
\end{longtable}

\cleardoublepage
\bibliography{sekharee}
\newpage
\section*{Graphical abstract}
\begin{figure}[h!]
	\centering
	\includegraphics[width=0.7\textwidth]{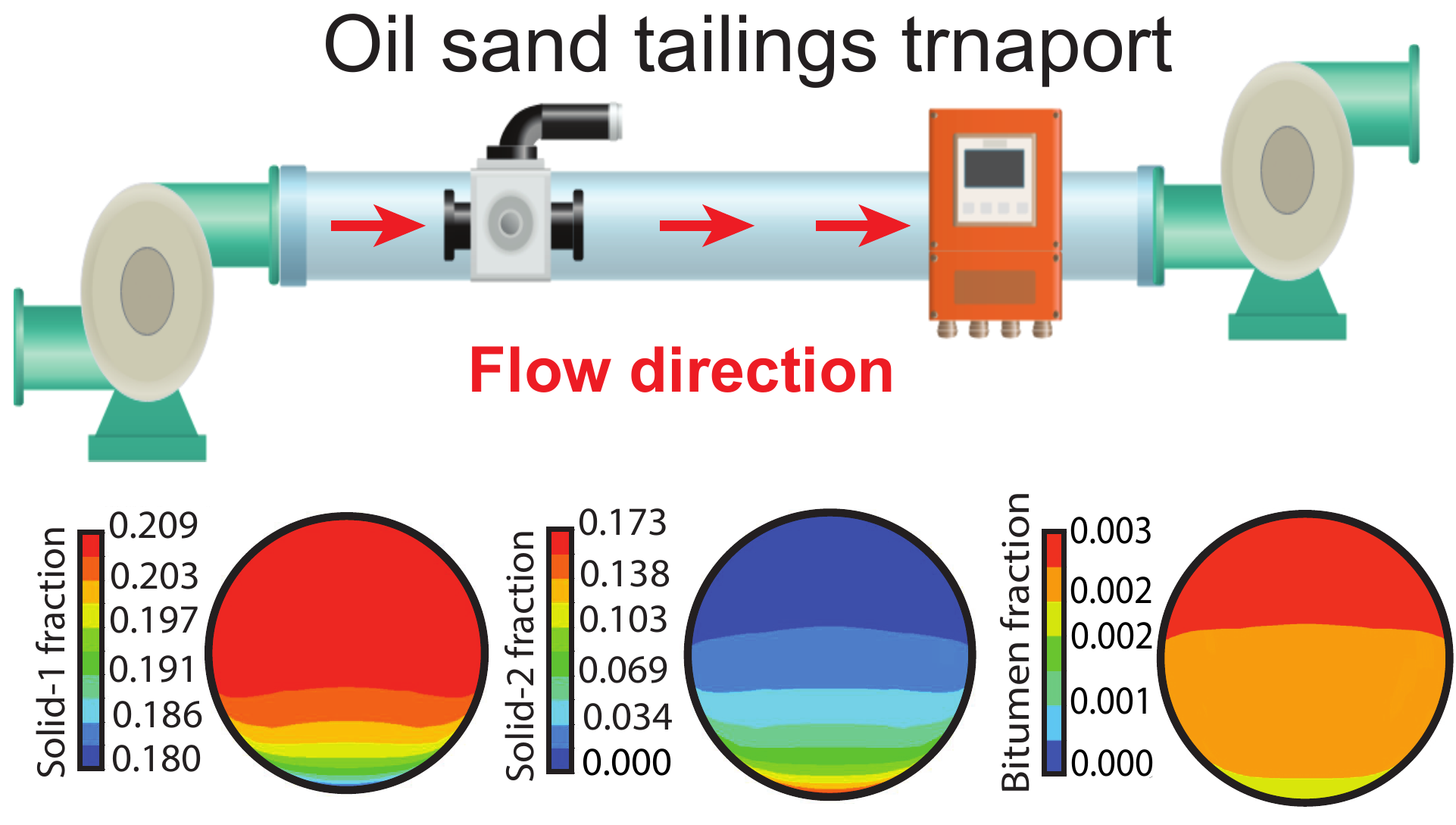}
\end{figure}
\section*{Highlights}

\begin{itemize}
	
 \item A 3D CFD model was developed for turbulent tailings slurries in a pipeline.

\item Excellent agreement between model validation and field data.

\item Bitumen recovery was analyzed in relation to bubble size and bubble fraction.
\item Revealed new insights into the distribution of bitumen in a horizontal industrial pipe. 
	
\end{itemize}

\end{document}